\documentclass[onefignum,onetabnum]{siamart190516}

\usepackage{lipsum}
\usepackage{amsfonts}
\usepackage{graphicx}
\usepackage{epstopdf}
\usepackage{algorithmic}
\ifpdf
  \DeclareGraphicsExtensions{.eps,.pdf,.png,.jpg}
\else
  \DeclareGraphicsExtensions{.eps}
\fi

\newsiamremark{remark}{Remark}
\newsiamremark{hypothesis}{Hypothesis}
\crefname{hypothesis}{Hypothesis}{Hypotheses}
\newsiamthm{claim}{Claim}

\headers{Quasi-steady conduction limited melting}{L. C. Morrow et al.}

\title{Moving boundary problems for quasi-steady conduction limited melting\thanks{Draft document as of July 4, 2019.
}}

\author{Liam C. Morrow\thanks{School of Mathematical Sciences, Queensland University of Technology, QLD, 4001, Australia ({\tt scott.mccue@qut.edu.au})}
\and John R. King\thanks{School of Mathematical Sciences, University of Nottingham, Nottingham, NG7 2RD, United Kingdom}
\and Timothy J. Moroney\footnotemark[2]
\and Scott W. McCue\footnotemark[2] }

\usepackage{amsopn}

\usepackage{amsmath,amsfonts,mathtools,amssymb}
\usepackage{hyperref}
\hypersetup{colorlinks = true, linkcolor = blue, citecolor = blue}
\usepackage{esdiff}
\usepackage{graphicx,epstopdf,tikz,pgfplots}
\usepackage{cleveref}
\renewcommand{\vec}[1]{\mathbf{#1}}
\usepackage{pgfplots}
\usepackage{caption,subcaption}
\usepackage{xfrac}

\begin{document}

\maketitle

\begin{abstract}
	The problem of melting a crystal dendrite is modelled as a quasi-steady Stefan problem.  By employing the Baiocchi transform, asymptotic results are derived in the limit that the crystal melts completely, extending previous results that hold for a special class of initial and boundary conditions.   These new results, together with predictions for whether the crystal pinches off and breaks into two, are supported by numerical calculations using the level set method. The effects of surface tension are subsequently considered, leading to a canonical problem for near-complete-melting which is studied in linear stability terms and then solved numerically.  Our study is motivated in part by experiments undertaken as part of the Isothermal Dendritic Growth Experiment, in which dendritic crystals of pivalic acid were melted in a microgravity environment: these crystals were found to be prolate spheroidal in shape, with an aspect ratio initially increasing with time then rather abruptly decreasing to unity. By including a kinetic undercooling-type boundary condition in addition to surface tension, our model suggests the aspect ratio of a melting crystal can reproduce the same non-monotonic behaviour as that which was observed experimentally.
\end{abstract}



\section{Introduction} \label{sec:intro}

While there is a variety of simple models to approximate the shape of a melting particle \cite{jensen17,kintea15}, the traditional approach from a mathematical perspective is to employ a Stefan problem, which involves the linear heat equation subject to appropriate boundary conditions on the solid-melt interface.  These moving boundary problems are well studied via rigorous analysis, asymptotic techniques, some exact solutions and numerical computation.  Almost all of the analytical progress has been made for one-dimensional problems or those with radial symmetry \cite{gupta,Kondratiuk2015,mccue4,Moore2017,soward}, although there have been successful studies in which the symmetry is broken \cite{kingrileywallman,mccue1,mccue3,velazquez}.  We continue this direction in the present study, focusing on the melting of an axially symmetric dendritic crystal.  We employ both analytical and numerical techniques to study the shape of the evolving crystal, focussing on the very final stages of melting.

A key aspect of a traditional Stefan problem is that the effects of convection are ignored.  An excellent example of a relevant physical application involves certain experiments undertaken on the space shuttle Columbia, as part of the so-called Isothermal Dendritic Growth Experiment (IDGE) \cite{glicksman0,glicksman2,glicksman}, in which convection is not an issue.  The conduction-limited melting that was studied in those experiments provides a physical motivation for the kind of theoretical Stefan problems considered here.  A brief summary of these experiments is as follows.  A pure liquid melt, pivalic acid, is held at a temperature $u^*>u_{\mathrm{m}}^*$, where $u_{\mathrm{m}}^*\approx 35.9$~$^\circ$C is the equilibrium melting temperature.  The temperature is then reduced to slightly supercool the melt so that $u\lesssim u_{\mathrm{m}}$ throughout.  The growth of dendrites is initiated by activating a thermoelectric cooler to chill a small isolated volume of the melt, leading to a dendritic mushy zone.  Finally, the temperature is raised to remelt the crystals, returning the system to a stable melt phase.

\begin{figure}
	\centering
	\includegraphics[width=0.7\linewidth]{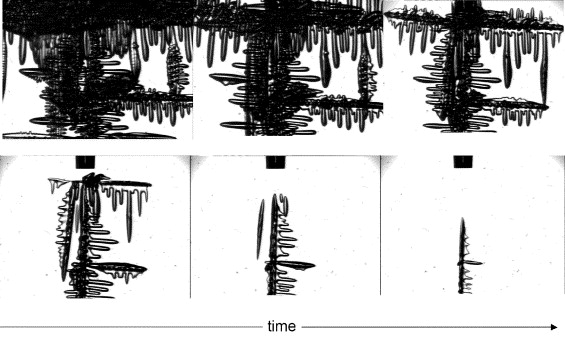}
	\caption{Left: A sequence of video frames of melting ellipsoidal PVA crystal collected as part of the Isothermic Dendritic Growth experiment. Right: Digital analysis of the middle frame on the left. The major, $C$, and minor, $A$, axis where computed using automated edge detection software to approximate the aspect ratio as a function of time. The black tip of the glass injector at the top of each frame has a diameter of 1~mm. Reproduced from Glicksman et al.~\cite{glicksman2} with permission from Springer Nature.}
	\label{fig:experiment}
\end{figure}

We are particularly interested in the final component of the IDGE.  After sufficient melting of the mushy zone had occurred, the remaining fragments consisted of isolated crystallites that resulted from partially melting dendritic side branches.  Typically these were roughly prolate spherical in shape (see \Cref{fig:experiment}).  For the final minute of melting of a particular crystal, video data (filmed at 30 frames per second) was analysed to determine the aspect ratio at each time.  For the examples presented by Glicksman and co-workers \cite{glicksman2,glicksman}, the aspect ratio of the needle-shaped crystals increased with time from about 7 at $t_\mathrm{e}-t=60$~s to 17 at $t_\mathrm{e}-t=10$~s, where $t_\mathrm{e}$ is the final melting time (also referred to as the extinction time).  After about $t_\mathrm{e}-t=10$~s, the aspect ratio rapidly decreased, and appeared to approach unity as $t\rightarrow t_\mathrm{e}^-$, meaning that the crystals were spherical just before extinction.

In order to make analytical progress, Glicksman et al. \cite{glicksman0} model the process with a one-phase quasi-steady problem, which results by ignoring heat conduction within the crystals and assuming an infinite Stefan number.  Here, the Stefan number is defined by
\begin{equation}
\beta=\frac{L}{c(u_{\infty}^*-u_{\mathrm{m}}^*)},
\label{eq:stefan}
\end{equation}
where $c$ is the specific heat, $L$ the latent heat of fusion per mass and $u_{\infty}^*-u_{\mathrm{m}}^*$ is the temperature difference between the melt away from the crystal and the melting temperature.  In reality, for this particular experiment the parameter values were $L/c\approx 10.99$~K, $u_{\infty}^*-u_{\mathrm{m}}^*\approx 1.8$~K, so $\beta\approx 6.1$, which is not reasonably large.  Glicksman et al. \cite{glicksman0} derive an exact solution to the infinite-Stefan-number problem in an infinite domain in prolate spheroidal coordinates, which applies under the further assumption that the aspect ratio of the dendrite remains constant.  This solution is a special case of that presented earlier by Ham~\cite{ham59} and Howison \cite{howison1} (which holds for the more general shape of an ellipsoid with constant aspect ratios), for example, and that derived using the Baiocchi transform by McCue et al.~\cite{mccue2} (again, for an ellipsoid).  The solution was used by Glicksman et al.~\cite{glicksman0} to approximate the time-dependence of the melting process, with quite good agreement with experimental results.

Glicksman and co-workers \cite{glicksman0,glicksman2,glicksman} did not provide an explanation for the observed increase in aspect ratio during the first 50~s of melting; however, the subsequent decrease in aspect ratio (during the final 10~s of melting) was accounted for by noting that by this stage of the melting process the crystals had become small enough for surface tension effects to begin to dominate~\cite{glicksman2,glicksman}.  As a consequence, the needle tips with high curvature melted more quickly than the remainder of the crystals, in accordance with the Gibbs-Thomson law
\begin{equation}
u^*=u_{\mathrm{m}}^*(1-\gamma\kappa^*)\quad\mbox{on}\quad\partial\Omega^*,    \label{eq:gibbsthoms}
\end{equation}
which states that the actual melting temperature on a curved surface is not constant, but instead depends weakly on the mean curvature $\kappa^*$ (defined to be positive for a sphere) via the surface tension coefficient $\gamma$ (defined to be $\gamma=2\sigma^*/\rho_s L$, where $\sigma^*$ measures surface energy effects with dimensions Nm$^{-1}$ or Jm$^{-2}$ and $\rho_s$ is the density of the solid phase)~\cite{back14b}.  Here $\partial\Omega^*$ denotes the solid-melt interface.  For the IDGE experiments, the surface tension coefficient is roughly $\gamma\sim 10^{-10}\,$~m.

In this article, we are motivated by these issues to undertake a theoretical study of the one-phase quasi-steady Stefan problem.  The mathematical problem is re-formulated in \Cref{sec:zerosurfacetension} with a Baiocchi transform for the special zero-surface-tension case.  In \Cref{sec:extinction}, we go on to provide a near extinction analysis for a general shaped initial crystal, including numerical results for cases in which crystals ultimately melt to a single point or pinch off and break into two separate pieces.  The role of surface tension is then explored in \Cref{sec:SurfaceTension}, while in \Cref{sec:KineticUndercooling} we consider an additional effect on the moving boundary, kinetic undercooling.  We show that kinetic undercooling acts as a de-stabilising term, and is effectively in competition with surface tension. When these two terms are considered simultaneously, we find that the aspect ratio of a prolate spheroid can initially increase before decreasing suddenly to unity in the extinction limit, which is the same behaviour as observed in the IDGE.  We close in \Cref{sec:discussion} with a summary of the key results and a brief discussion of how our work relates to the experiments described by Glicksman and co-workers~\cite{glicksman0,glicksman2,glicksman}.  An important point to note is that the quasi-steady assumption used in this article leads to a moving boundary problem that also describes bubble contraction in a porous medium \cite{dibenedetto,howison1,mccue2}.  Thus our study also describes the effect that surface tension has on the shape of a bubble in the limit that it contracts to a point.  This connection is revisited in \Cref{sec:discussion}.

\section{Quasi-steady formulation with zero surface tension}\label{sec:zerosurfacetension}

\subsection{Governing equations}\label{sec:goveqns}

Consider a solid substance (the crystal dendrite), initially at melting temperature $u_{\mathrm{m}}^*$ occupying the region $\Omega^*(0)$, surrounded by the same substance in liquid form in $\mathbb{R}^{3}\setminus\Omega^*$.  In the far field, a higher temperature $u_{\infty}^*$ is applied, and thus melting proceeds until the crystal melts completely at the extinction time $t_\mathrm{e}^*$.

Setting $k$ to be the thermal diffusivity, we scale variables using
\begin{equation}
t=\frac{k}{\ell^2\beta}t^*,
\quad
{\bf x}=\frac{1}{\ell}{\bf x}^*,
\quad
u=\frac{u^*-u_{\mathrm{m}}^*}{u_{\infty}^*-u_{\mathrm{m}}^*},
\end{equation}
where $\ell$ is a characteristic length scale of the initial crystal shape, and $\beta$ is the Stefan number (\ref{eq:stefan}).  The resulting one-phase Stefan problem for melting the crystal is
\begin{subequations}
	\begin{align}
	& \mbox{in}\quad \mathbb{R}^3\setminus\Omega(t): && \frac{1}{\beta}\frac{\partial u}{\partial t}=\nabla^2 u,
	\label{eq:heat_u}\\
	& \mbox{on}\quad\partial\Omega: && u=0,
	\label{eq:zerostcondition}\\
	& \mbox{on}\quad\partial\Omega: && V_n=-\frac{\partial u}{\partial n}, \label{eq:kinematicboundary} \\
	& \mbox{as}\quad r\rightarrow\infty: &&  u \rightarrow 1,
	\label{eq:outerboundary}
	\end{align}
	where $V_n$ represents the normal velocity of the solid-melt interface $\partial\Omega$, defined to be negative for a shrinking surface.
	
	For what follows we shall take the quasi-steady limit $\beta=\infty$, which is an appropriate approximation for experiments in which the latent heat is large or the specific heat is small.  As a result, the parabolic equation \eqref{eq:heat_u} becomes Laplace's equation
	\begin{align}
	& \mbox{in}\quad \mathbb{R}^3\setminus\Omega(t): \qquad \nabla^2 u=0,		\label{eq:Laplace}
	\end{align}
\end{subequations}
and thus we do not require an initial condition for $u$.

As mentioned in the Introduction, the governing equations \eqref{eq:Laplace} with \eqref{eq:zerostcondition}-\eqref{eq:outerboundary} are also relevant for the problem of a bubble that is forced to contract in a saturated medium, where the fluid flow is governed by Darcy's law \cite{dibenedetto,howison1,mccue2}, as well as the two-dimensional analogue for Hele-Shaw flow \cite{entov,entov2,lee}.  These equations also arise in other moving boundary problems, for example the small P{\'e}clet number limit of advection-diffusion-limited dissolution/melting models \cite{Claudin2017,Hewett2017,Mac2015,Rycroft2016,Wykes2018}, for which it is also of interest to track the moving boundary and predict its shape and location (the collapse point \cite{Rycroft2016}) close to the extinction time; other closely related advection-diffusion-like moving boundary problems in potential flow have similar governing equations in the small P{\'e}clet number limit \cite{bazant06,cummings99}.

\subsection{Baiocchi transform}\label{sec:baiocchi}

We use the Baiocchi transform defined by
\begin{subequations}
	\begin{align}
	& \mbox{in}\quad \mathbb{R}^3\setminus\Omega(0): && w=\int_0^t u({\bf x},t')\,\mathrm{d}t'\\
	& \mbox{in}\quad \Omega(0)\setminus\Omega(t): && w=\int_{\omega({\bf x})}^t u({\bf x},t')\,\mathrm{d}t',
	\end{align}
\end{subequations}
where we are using the notation $t=\omega({\bf x})$ to denote the solid-melt interface $\partial\Omega$.  The Baiocchi transform is widely used in the analysis of moving boundary problems with boundary conditions of the form (\ref{eq:zerostcondition})-(\ref{eq:kinematicboundary}), for example \cite{cummings99b,Elliott81,howisonking,kingmccue,Lacey1982,mccue2}.  Note that while here we restrict ourselves to (\ref{eq:Laplace}), the approach is also applicable to (\ref{eq:heat_u}) \cite{mccue1,mccue3}.

Transforming the governing equations \eqref{eq:Laplace} with \eqref{eq:zerostcondition}-\eqref{eq:outerboundary}, we derive the nonlinear moving boundary problem for $w$:
\begin{subequations}
	\begin{align}
	& \mbox{in}\quad \mathbb{R}^3\setminus\Omega(0): && \nabla^2 w=0,
	\label{eq:baiocchi6}\\
	& \mbox{in}\quad \Omega(0)\setminus\Omega(t): && \nabla^2 w=1,
	\label{eq:baiocchi7}\\
	& \mbox{on}\quad\partial\Omega: && w=0,
	\label{eq:baiocchi8}\\
	& \mbox{on}\quad\partial\Omega: && \frac{\partial w}{\partial n}=0,
	\label{eq:baiocchi9}\\
	& \mbox{as}\quad r\rightarrow\infty: &&  w\rightarrow t.
	\label{eq:baiocchi10}
	\end{align}
\end{subequations}
Once a solution for the Baiocchi variable $w$ is determined, the temperature $u$ can be recovered via $u=\partial w/\partial t$.  We note that an advantage of the Baiocchi transform is that it transforms the inhomogeneous boundary condition \eqref{eq:kinematicboundary} into a homogeneous boundary condition.  Another is that time appears as a parameter in \eqref{eq:baiocchi6}-\eqref{eq:baiocchi10}, so that the problem can be solved at any time without knowledge of the solution at previous times.

\subsection{Exact solution for prolate spheroid} \label{sec:exact}

For the case in which the initial crystal shape $\partial\Omega(0)$ is an ellipsoid,  \eqref{eq:baiocchi6}-\eqref{eq:baiocchi10} can be solved in ellipsoidal coordinates exactly, as done as part of the analysis by McCue et al.~\cite{mccue2}.  The solution for the interface $\partial\Omega(t)$ remains ellipsoidal with constant aspect ratios for all time until extinction.  An equivalent solution without the Baiocchi transform is provided in Howison \cite{howison1}.

We present here a summary of this exact solution in the special case for which the initial crystal shape $\partial\Omega(0)$ is the prolate spheroid
\begin{equation}
x^2+y^2+\frac{z^{2}}{z_0(0)^2}=1,	\label{eq:prolateinit}
\end{equation}
with initial aspect ratio $\mathcal{A}(0)=z_0(0)$.  (This special case, together with the case in which the crystal is initially an oblate spheroid, is also recorded by McCue et al.~\cite{mccue2}.)
The exact solution is that $\partial\Omega(t)$ retains its prolate spheroidal shape as
\begin{equation}
\frac{x^2+y^2}{\rho_0(t)^2}+\frac{z^2}{z_0(t)^2}=1,    \label{eq:prolate}
\end{equation}
where $z_0(t)>0$ and $\rho_0(t)>0$ measure the major and minor axes of the dendrite, respectively, with constant aspect ratio $\mathcal{A}(t)=z_0(t)/\rho_0(t)=z_0(0)$ (here the length scale $\ell$ is chosen so that $\rho_0(0)=1$).  The full solution has the time-dependence
\begin{equation}
\frac{z_0(t)}{z_0(0)}=\rho_0(t)=\sqrt{1-\frac{t}{t_\mathrm{e}}},
\label{timedep}
\end{equation}
where
\begin{equation}
t_\mathrm{e}=\frac{z_0(0)}{4\sqrt{z_0(0)^2-1}}\ln\left(
\frac{z_0(0)+\sqrt{z_0(0)^2-1}}{z_0(0)-\sqrt{z_0(0)^2-1}}
\right).
\label{exactextin}
\end{equation}
The result \eqref{exactextin} is also derived in Glicksman et al.~\cite{glicksman0}.
Although, as mentioned above, the aspect ratio of the melting crystals in the Isothermal Dendritic Growth Experiment was not constant, these authors make a rough guess for the average value of the aspect ratio over the first 50 seconds of melting, and then compare  \eqref{timedep} with experimental results.  Their agreement is quite good, reflecting the square root of time dependence near extinction.

\section{Analysis of zero-surface-tension problem}\label{sec:extinction}

McCue et al.~\cite{mccue2} were concerned primarily with analysing the near extinction behaviour for a variation of  \eqref{eq:baiocchi6}-\eqref{eq:baiocchi10} in which $\Omega(0)$ coincides with an outer boundary (i.e., a finite-domain problem in which the crystal initially occupies the entire domain).  Here we provide equivalent results for the full infinite-domain problem \eqref{eq:baiocchi6}-\eqref{eq:baiocchi10} and apply the level set method to support these findings.

\subsection{Extinction time and extinction points}\label{sec:extincttimepoint}

For a given initial crystal shape $\Omega(0)$, we wish to determine how long it takes to melt (the extinction time $t_\mathrm{e}$) and the point at which the crystal eventually vanishes as $t\rightarrow t_\mathrm{e}^-$ (the extinction point ${\bf x}_\mathrm{e}$).  The convenient framework for this analysis is via the Baiocchi transform.  As mentioned above, time appears as a parameter in \eqref{eq:baiocchi6}-\eqref{eq:baiocchi10}, meaning we can skip to the extinction time to compute $w_\mathrm{e}({\bf x})=w({\bf x},t_\mathrm{e})$.  It is convenient to set $w_\mathrm{e}=W({\bf x})+t_\mathrm{e}$, so $W$ satisfies the linear problem
\begin{subequations}
	\begin{align}
	& \mbox{in}\quad \mathbb{R}^3\setminus\Omega(0): && \nabla^2 W=0, 	\label{eq:Baiocchi1}\\
	& \mbox{in}\quad \Omega(0): && \nabla^2 W=1, 						\label{eq:Baiocchi2}\\
	& \mbox{as}\quad r\rightarrow\infty: &&  W\rightarrow 0. 			\label{eq:Baiocchi3}
	\end{align}
\label{eq:Baiocchi}
\end{subequations}
The extinction point ${\bf x}_\mathrm{e}$ is then the local maximum of $W$, and the extinction time is recovered via $t_\mathrm{e}=-W({\bf x}_\mathrm{e})$.  As noted by Entov \& Etingof~\cite{entov}, \eqref{eq:Baiocchi1}-\eqref{eq:Baiocchi3} defines the dimensionless gravity potential of $\Omega(0)$, thus
\begin{equation}
W=-\frac{1}{4\pi}\int\!\!\!\int\!\!\!\int_{\Omega(0)}\frac{1}{|{\bf x}-{\bf x}'|}\,\mathrm{d}V',
\label{eq:gravitypotential}
\end{equation}
which provides an interesting connection between our problem and gravity potential generated by a uniform body.

Whilst in practice it is not feasible to compute $W$ analytically for a general initial crystal shape $\Omega(0)$, such a calculation can be performed numerically.  Indeed, we provide a number of simple examples in \Cref{sec:numericalzst} in which we compute $W$ for both convex and non-convex initial shapes.  We include in those examples cases for which $W$ has two local maxima.  In such instances, if the two local maxima are equal, then the crystal must pinch off into two, with the local maxima corresponding to the extinction points for each of the two satellite crystals.  We also provide an example of the more complicated case in which there are two local maxima that are not equal; here, the use of $W$ can only predict the final extinction for the largest of the two satellite crystals.

\subsection{Near-extinction analysis}\label{sec:nearextinct}

For the case of an axially symmetric initial crystal with the $z$ axis pointing down the centreline, we can translate the coordinate system so that the extinction point ${\bf x}_\mathrm{e}$ lies on the origin.  Since $w_\mathrm{e}=0$ at ${\bf x}={\bf x}_\mathrm{e}$ and ${\bf x}_\mathrm{e}$ is a local maximum of $w_\mathrm{e}$, a simple Taylor series for this axially symmetric geometry implies that $w_\mathrm{e}\sim a(x^2+y^2)+bz^2$ as $r\rightarrow 0$.  Further, as a consequence of \eqref{eq:Baiocchi2}, we then have
\begin{equation}
w_\mathrm{e}\sim a(x^2+y^2)+\left(\frac{1}{2}-2a\right)z^2
\quad\mbox{as}\quad r\rightarrow 0,
\label{eq:smallrlimit}
\end{equation}
where $1/6<a<1/4$. As we shall see, the parameter $a$ is effectively all the melting crystal ``remembers'' from its initial condition; it is this single parameter that controls the aspect ratio of the crystal at extinction.  Note that the higher order terms in (\ref{eq:smallrlimit}) are not required in the following analysis (they would be for the special case $a=1/4$, which represents the borderline between the type of extinction considered in this section and when a bubble breaks up into two, as treated in \Cref{sec:numericalzst}).

In the limit $t\rightarrow t_\mathrm{e}^-$, the inner region is for $r=\mathcal{O}(T)$, where $T(t)$ is a length scale defined so that the volume of the melting crystal is fixed to be $4\pi T^3/3$.  We write $w\sim T^2 \Phi({\bf X})$ as $t\rightarrow t_\mathrm{e}^-$, where ${\bf X}={\bf x}/T$, so that
\begin{subequations}
	\begin{align}
	& \mbox{in}\quad \mathbb{R}^3\setminus\Omega_0(0): && \frac{\partial^2\Phi}{\partial X^2}
	+\frac{\partial^2\Phi}{\partial Y^2}+\frac{\partial^2\Phi}{\partial Z^2}=1,
	\label{eq:baiocchi11}\\
	& \mbox{on}\quad\partial\Omega_0: && \Phi=0, \quad \frac{\partial \Phi}{\partial N}=0,
	\label{eq:baiocchi12}
	\end{align}
	where $\Omega_0$ denotes the crystal which has volume $4\pi/3$ in these self-similar coordinates, and $N$ denotes a normal direction.  In order to match with \eqref{eq:smallrlimit} we require that \begin{equation}
	\Phi\sim  a(X^2+Y^2)+\left(\frac{1}{2}-2a\right)Z^2-d+\frac{1}{3R},
	\label{eq:baiocchi13}
	\end{equation}
\end{subequations}
as $R\rightarrow\infty$, where $d$ is a constant found as part of the solution to \eqref{eq:baiocchi11}-\eqref{eq:baiocchi13}.  We see from \eqref{eq:baiocchi13} that a matching condition for the outer region is
\begin{equation}
w\sim a(x^2+y^2)+\left(\frac{1}{2}-2a\right)z^2-dT^2+\frac{T^3}{3r}
\quad\mbox{as}\quad r\rightarrow 0.
\end{equation}

The solution to \eqref{eq:baiocchi11}-\eqref{eq:baiocchi13} in prolate spheroidal coordinates is provided in \Cref{sec:appendix}.  According to this solution the dendrite boundary $\partial\Omega_0$ is described by
\begin{equation}
\frac{X^2+Y^2}{q_0^2-1}+\frac{Z^2}{q_0^2}=\frac{1}{q_0^{2/3}(q_0^2-1)^{2/3}},
\end{equation}
where $q_0$ is a parameter that is related to the special constant $a$ by
\begin{equation}
a=\frac{1}{4}q_0^2-\frac{1}{8}q_0(q_0^2-1)\ln\left(\frac{q_0+1}{q_0-1}\right).
\label{eq:a}
\end{equation}
Further, the constant $d$ in \eqref{eq:baiocchi13} is related implicitly to $a$ by
\begin{equation}
d=\frac{1}{4}q_0^{1/3}(q_0^2-1)^{1/3}\ln\left(\frac{q_0+1}{q_0-1}\right).
\label{eq:d}
\end{equation}
Note that the prolate spheroid approaches a perfect sphere in the limit $a\rightarrow 1/6^+$, in which case $d\rightarrow 1/2^+$.

The outer region is for $r=\mathcal{O}(1)$, for which
\begin{equation}
w\sim w_\mathrm{e}-(t-t_\mathrm{e})+\frac{T^2}{3r}
\quad\mbox{as}\quad t\rightarrow t_\mathrm{e}^-.
\end{equation}
Matching with the inner gives the time-dependence
\begin{equation}
t= t_\mathrm{e}-dT^2+\mathcal{O}(T^5)\quad\mbox{as}\quad T\rightarrow 0,
\end{equation}
or, in other words,
\begin{equation}
T\sim\frac{1}{\sqrt{d}}(t_\mathrm{e}-t)^{1/2}
\quad\mbox{as}\quad t\rightarrow t_\mathrm{e}^-.
\label{eq:timedependence}
\end{equation}
Thus we see that, regardless of the shape of the initial crystal, the square root of time scaling determined experimentally in Glicksman et al.~\cite{glicksman0} is as expected.

In summary, the zero-surface-tension model predicts that, provided there is no pinch-off, an axially symmetric dendrite will melt to a spheroid in the extinction limit.  While this spheroid could be prolate or oblate, we concentrate here on the prolate case, as this is the one observed in the IDGE \cite{glicksman0,glicksman2,glicksman}.  The aspect ratio of the prolate spheroid at extinction is given by
\begin{equation}
\mathcal{A}(t_\mathrm{e})=\frac{q_0}{\sqrt{q_0^2-1}},
\label{eq:aspectratio}
\end{equation}
which provides an implicit dependence of $\mathcal{A}$ on the constant $a$ via \eqref{eq:a}.  Here $a$ is the only parameter that is required to characterise the initial dendrite shape (it is found by solving \eqref{eq:gravitypotential} and expanding $w_\mathrm{e}$ about ${\bf x}_\mathrm{e}$).  The time-dependence of the melting is given by \eqref{eq:timedependence}, where the volume of the dendrite shrinks like $4\pi T^3/3$ (in other words, $T$ provides a natural length scale for the melting dendrite).  Again, this time-dependence is related to the initial dendrite shape via the parameter $a$ (since $d$ is given by $a$ through \eqref{eq:d} and \eqref{eq:a}).

In the special case in which the dendrite is initially the prolate spheroid \eqref{eq:prolateinit}, then it retains its aspect ratio.  This is, of course, the exact solution listed in \Cref{sec:exact}.

Finally, for sufficiently symmetric crystals we have $a=1/6$ which gives $d=1/2$.  Here $\Phi=R^2/2-1/2+1/3R$ and the dendrite becomes spherical in the limit with $T\sim\sqrt{2}(t_\mathrm{e}-t)^{1/2}$.  The special case of an initially spherical dendrite remains spherical.

At this point it is worth mentioning that for large Stefan numbers, $\beta\gg 1$, the scaling (\ref{eq:timedependence}) eventually ceases to hold for the full classical Stefan problem with (\ref{eq:heat_u}) instead of (\ref{eq:Laplace}) \cite{mccue3}. However, this discrepancy would not be observed on the scale of the IDGE experiments.

\subsection{Null quadrature domains}\label{sec:nullquad}

It is worth relating some of the above arguments to well-known and long-established results \cite{dibenedetto,friedman,howison1}.  First, by applying Green's theorem it can be shown that
\begin{equation}
\frac{\mathrm{d}}{\mathrm{d}t}\int\!\!\!\int\!\!\!\int_{\mathbb{R}^3\setminus\Omega(t)}\!\!\!\Phi({\bf x})\,\mathrm{d}V=0,
\label{eq:ratemoment3D}
\end{equation}
where $\Phi$ is a suitable harmonic function and $\Omega(t)$ is the shape of a melting crystal from the infinite-domain problem \eqref{eq:baiocchi6}-\eqref{eq:baiocchi10} (Howison~\cite{howison1}).  Noting that these quasi-steady problems with zero surface tension are time-reversible, we can seek so-called `ancient' solutions for which the entire domain $\mathbb{R}^3\setminus\Omega(t)$ vanishes in the limit $t\rightarrow -\infty$.  From \eqref{eq:ratemoment3D} it follows that for these ancient solutions $\mathbb{R}^3\setminus\Omega(t)$ must be a null quadrature domain.  The only suitable such domain is the exterior of an ellipsoid (see Karp~\cite{karp} for a discussion on null quadrature domains).  For any other initial crystal shape, the backwards problem with $t$ decreasing leads to some kind of finite-time blow-up or perhaps a scenario in which part of the crystal boundary expands infinitely leaving behind `fjords' or `tongues' (these scenarios are much better understood in the two-dimensional Hele-Shaw problem; see also Howison~\cite{howison2,howison3} for explicit examples of each case).

As discussed in \Cref{sec:extinction}, for a melting crystal (of general initial shape) the generic limiting behaviour is that it becomes ellipsoidal in shape as $t\rightarrow t_\mathrm{e}^-$.   This result can also be derived using an alternative approach, as suggested more recently by King \& McCue~\cite{kingmccue}, who treated the two-dimensional Hele-Shaw case.  First, we see that for the integral in \eqref{eq:ratemoment3D} to converge we could choose $\Phi= r^\ell Y_\ell^m$,
where $Y_\ell^m$ are spherical harmonics and $\ell$ is an integer such that $\ell\leq -4$. Rescaling lengths such that $\bar{r}=r/T$, we have from \eqref{eq:ratemoment3D} that
\begin{equation}
\int\!\!\!\int\!\!\!\int_{\mathbb{R}^3\setminus\bar{\Omega}(t)}\!\!\!\Phi(\bar{\bf x})\,\mathrm{d}\bar{V}
=\mathcal{O}(T^{-\ell-3})
\hspace{1.5ex} \mbox{as}\hspace{1.5ex} T\rightarrow 0
\hspace{1.5ex} \mbox{for} \hspace{1.5ex} \ell\leq -4.
\end{equation}
Thus, the left-hand side vanishes as $T\rightarrow 0$, or $t\rightarrow t_\mathrm{e}^-$, meaning that the exterior of the crystal approaches a null quadrature domain in the limit, and thus the crystal itself approaches an ellipsoid in shape.

\subsection{Numerical examples}\label{sec:numericalzst}

We present some numerical examples that demonstrate the key features discussed above.  To solve \eqref{eq:Baiocchi1}-\eqref{eq:Baiocchi3} numerically, we formulate a level set function, $\phi(\vec{x})$, such that $\phi>0$ for $\vec{x} \in \Omega(0)$ and $\phi<0$ for $\vec{x} \in \mathbb{R}^3 \backslash \Omega(0)$. Thus we can reformulate \eqref{eq:Baiocchi1} and \eqref{eq:Baiocchi2} as
\begin{equation} \label{eq:Baiocchi4}
\nabla^2 W = H(\phi),
\end{equation}
where $H$ is the Heaviside function. We note that $H(\phi)$ is discontinuous at $\vec{x} \in \partial \Omega(0)$, so for numerical purposes we implement a smoothed Heaviside function
\begin{equation}
\hat{H}(\phi) =
\begin{cases}
0 & \text{if } \phi < -\delta, \\
\frac{1}{2} \left( 1 + \frac{\phi}{\delta} + \frac{1}{\pi} \sin \frac{\pi \phi}{\delta} \right)  & \text{if } |\phi| \le \delta, \\
1 & \text{if } \phi > \delta,
\end{cases}
\end{equation}
where $\delta = 1.5 \Delta x$. For this purpose, it is convenient to work in spherical polar coordinates $(r,\theta,\varphi)$ and represent the axially symmetric moving boundary $\partial\Omega$ by $r=s(\theta,t)$. Thus, \eqref{eq:Baiocchi4} becomes
\begin{align} \label{eq:Baiocchi5}
\frac{1}{r^2} \diffp{}{r} \left( r^2 \diffp{W}{r} \right) + \frac{1}{r^2 \sin \theta} \diffp{}{\theta} \left( \sin\theta \diffp{W}{\theta} \right) = \hat{H}(\phi).
\end{align}
The spatial derivatives in  \eqref{eq:Baiocchi5} are approximated using central finite differencing, with homogeneous Neumann boundary conditions applied at $r = 0$, $\theta = 0$, and $\theta = \pi$. The far-field boundary condition \eqref{eq:Baiocchi3} is incorporated using a Dirichlet-to-Neumann map described in \Cref{sec:Farfield}.

\subsubsection{Symmetric initial condition} \label{sec:SymmetricIC}

We consider a selection of initial conditions to illustrate a few different qualitative behaviours.  Again, using spherical polar coordinates $(r,\theta,\varphi)$ with $\partial\Omega$ denoted by $r=s(\theta,t)$, the first is the prolate spheroid
\begin{equation}
s(\theta,0)=\frac{r_0}{\sqrt{r_0^2 \cos^2\theta + \sin^2 \theta}},		\label{eq:example1}
\end{equation}
where $r_0$ describes the initial aspect ratio.  The second initial condition is a peanut-shaped interface described by
\begin{equation}
s(\theta,0)=r_0+(1-r_0)\cos^2\theta,		\label{eq:examples2to4}
\end{equation}
where $r_0$ can be interpreted as a measure of the depth of the pinch in the middle of the peanut. Following Garzon et al.~\cite{Garzon2011}, the third initial condition considered is a dumbbell shaped interface of the form $s(\theta,0)=(\rho^*(\theta)^2+z^*(\theta)^2)^{1/2}$, where
\begin{subequations}
	\begin{align}
	z^*(\theta) &= 1 + r_0\sin^2(\theta / 2), 		\label{eq:SymmetricIC1a} \\
	\rho^*(\theta) &= g(\theta) + 2 g(\pi - \theta), 	\label{eq:SymmetricIC1b}
	\end{align}
	with
	\begin{align}
	g(\theta) &= \sqrt{r_0 k(\theta)} \left( \mathrm{e}^{-(r_0^2 k(\theta)^2)/2} - \mathrm{e}^{-r_0^2/2} \right),
	\label{eq:SymmetricIC1d} \\
	k(\theta) &= \cos^2 (\theta / 2),				\label{eq:SymmetricIC1c}
	\end{align}
\end{subequations}
for $0 \le \theta \le \pi/2$; for $\pi<\theta\leq 2\pi$ this initial condition is made symmetric by reflecting about $\theta=\pi/2$.

\begin{figure}
	\centering
	\begin{center} \eqref{eq:example1} with $r_0 = 0.8$ \end{center}
	\includegraphics[width=0.3\linewidth]{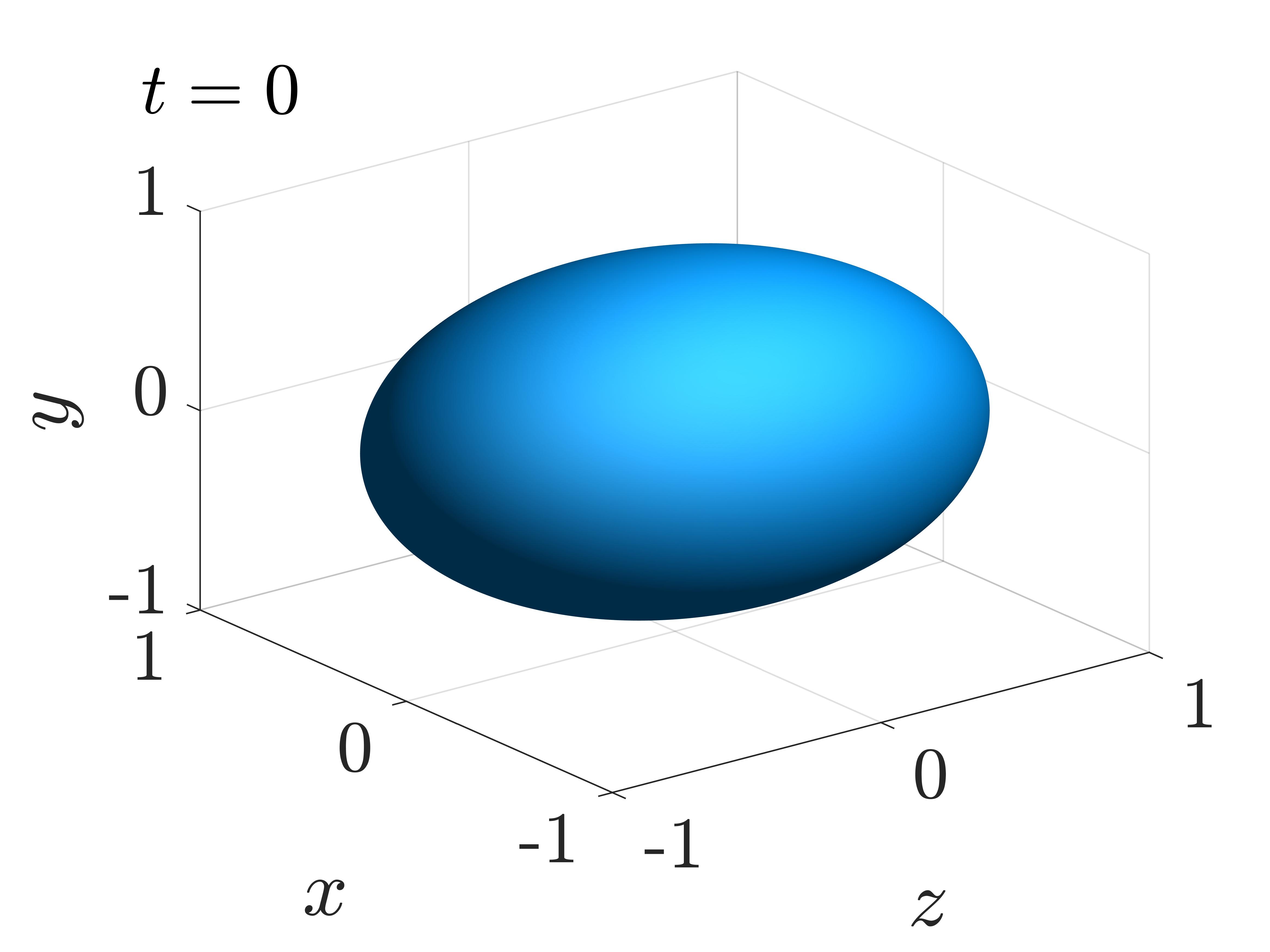}
	\includegraphics[width=0.3\linewidth]{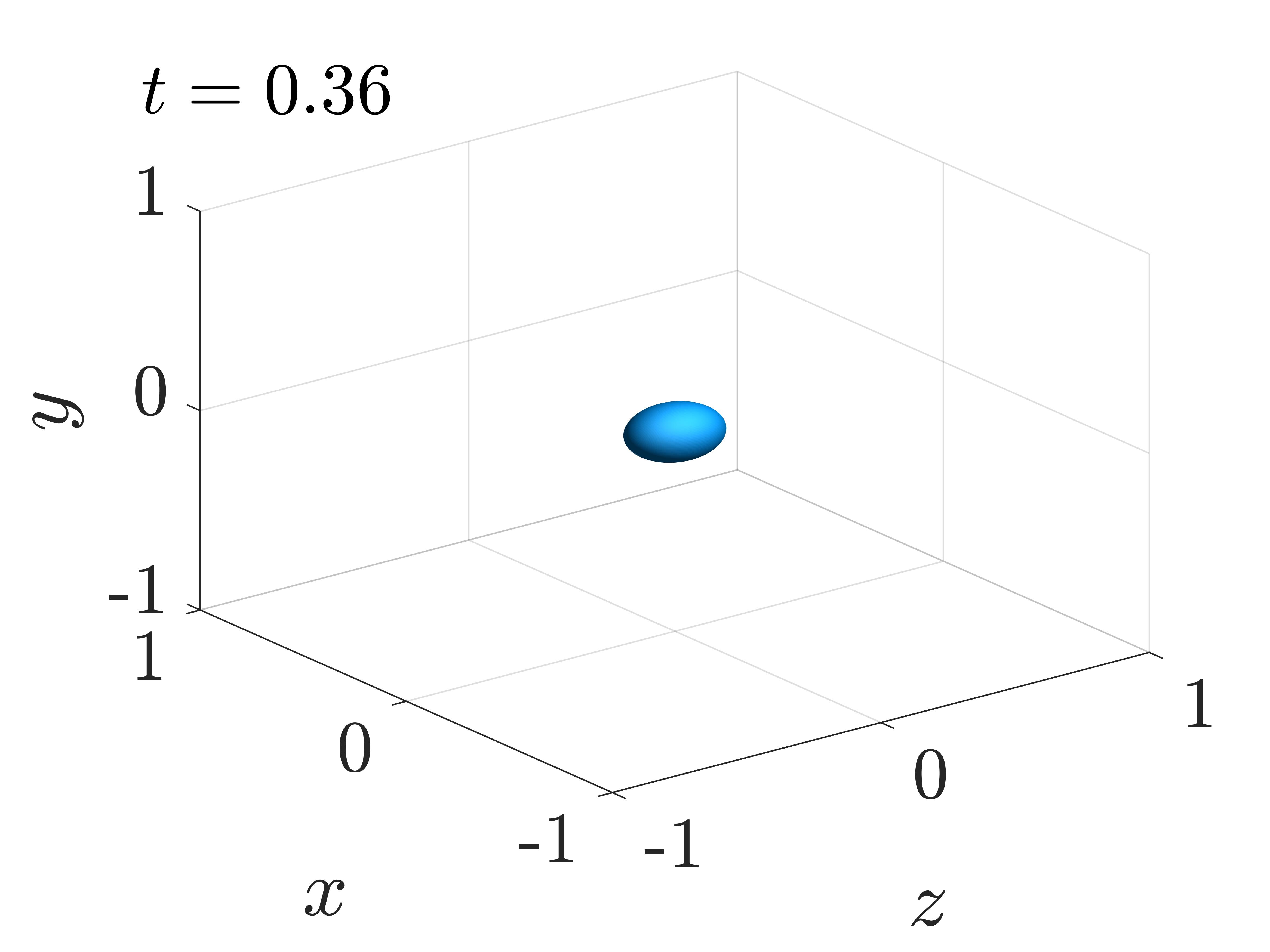}
	\includegraphics[width=0.3\linewidth]{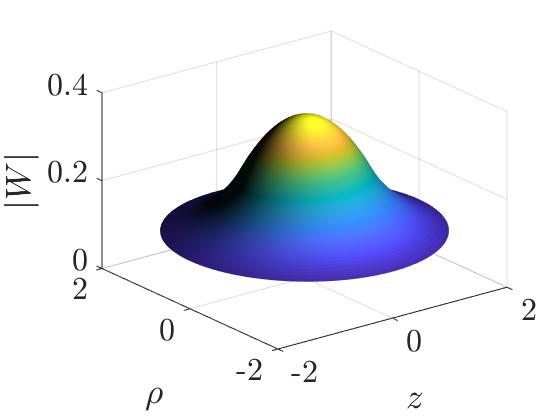}
	
	\begin{center} \eqref{eq:examples2to4} with $r_0 = 0.5$ \end{center}
	\includegraphics[width=0.3\linewidth]{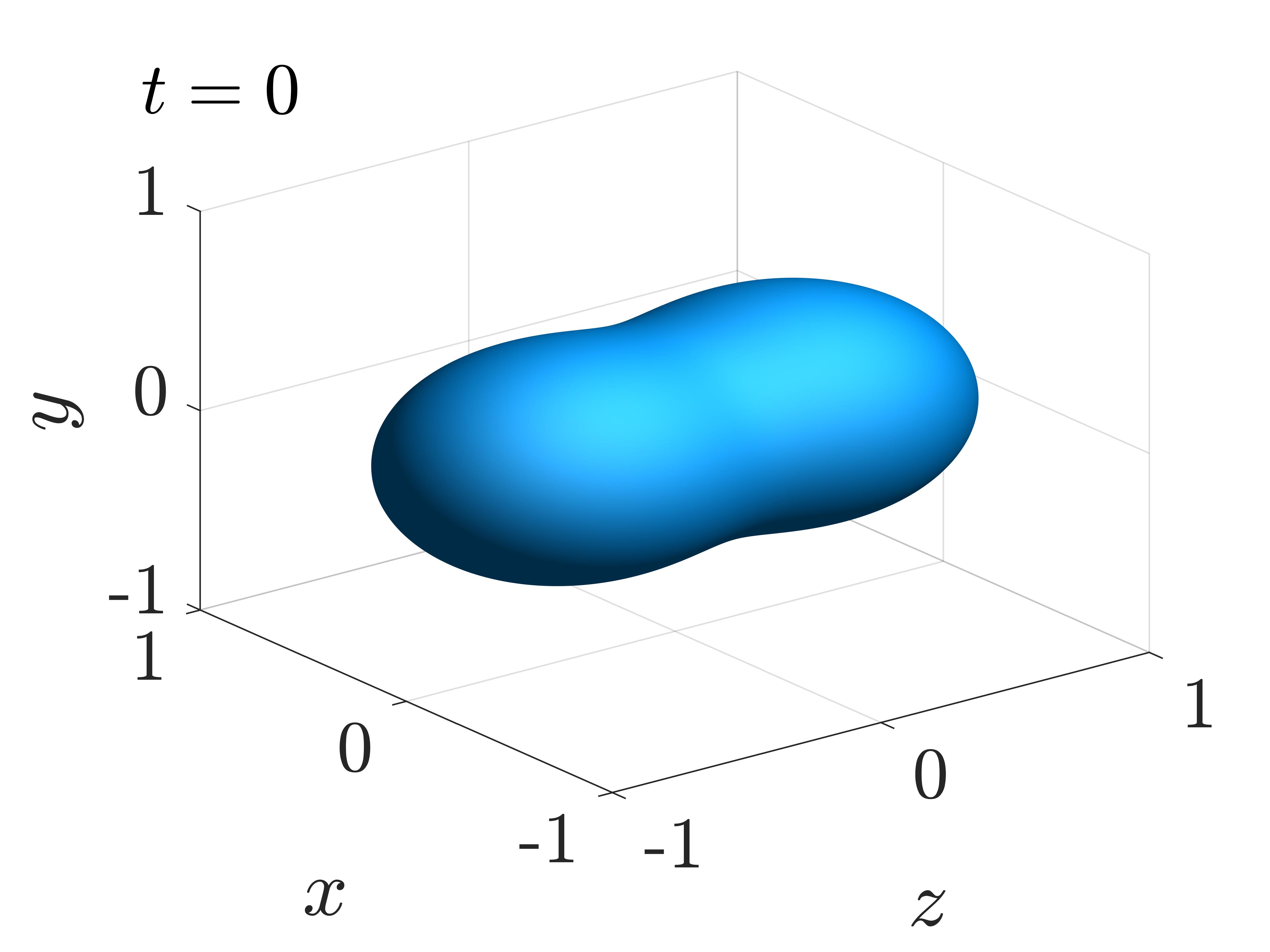}
	\includegraphics[width=0.3\linewidth]{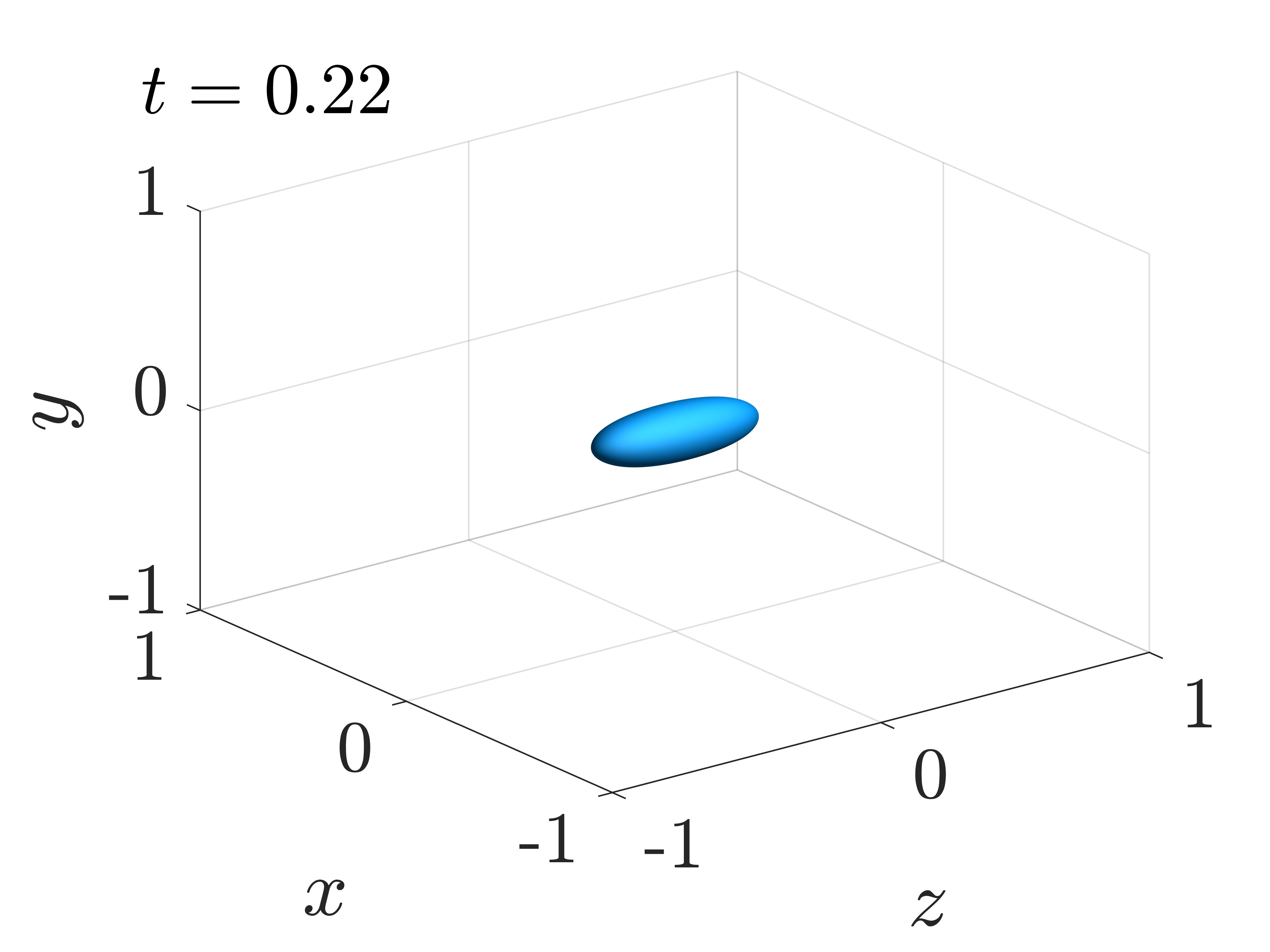}
	\includegraphics[width=0.3\linewidth]{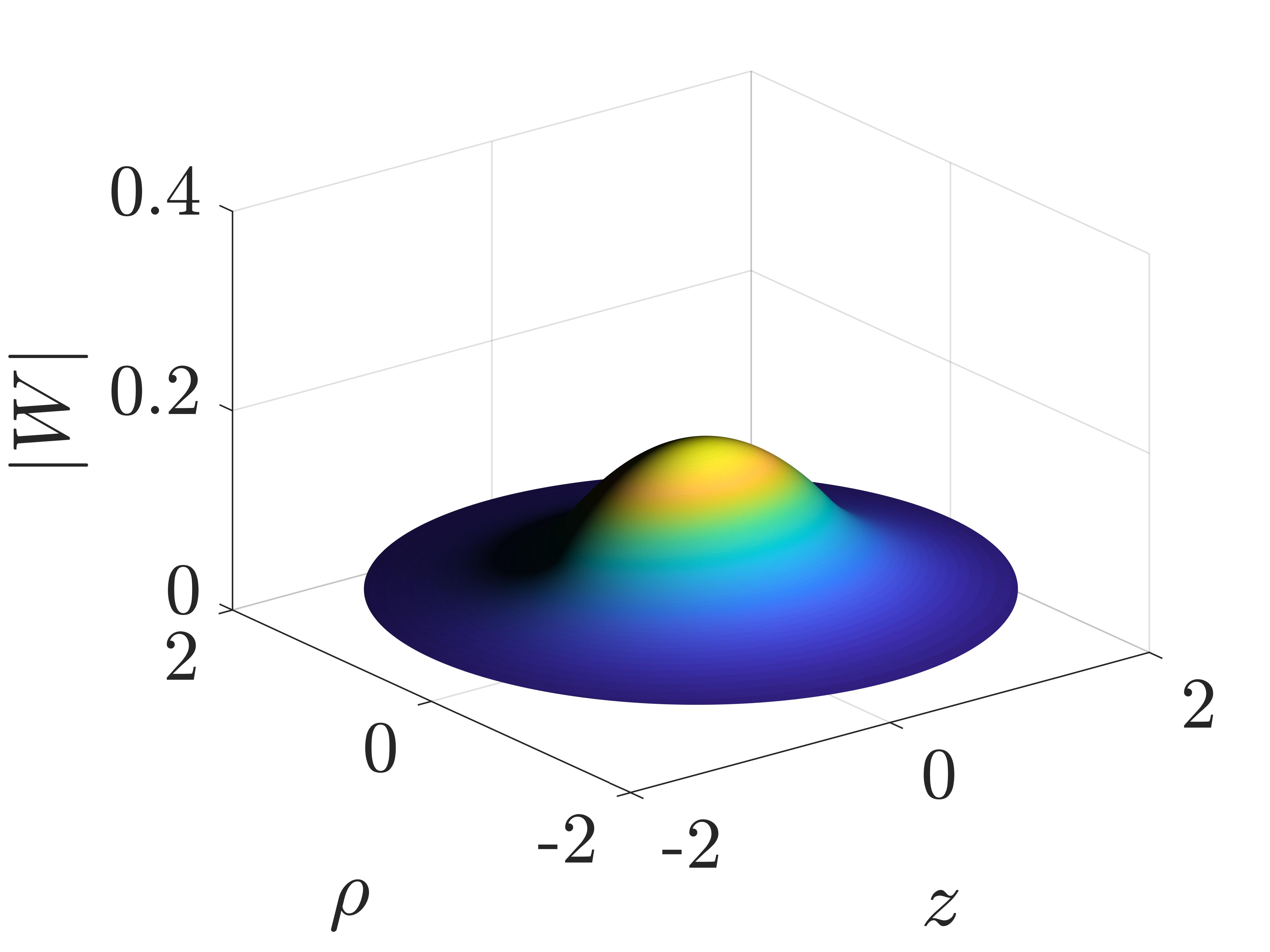}
		
	\begin{center} \eqref{eq:SymmetricIC1a}-\eqref{eq:SymmetricIC1c} with $r_0 = 4.75$  \end{center}
	\includegraphics[width=0.3\linewidth]{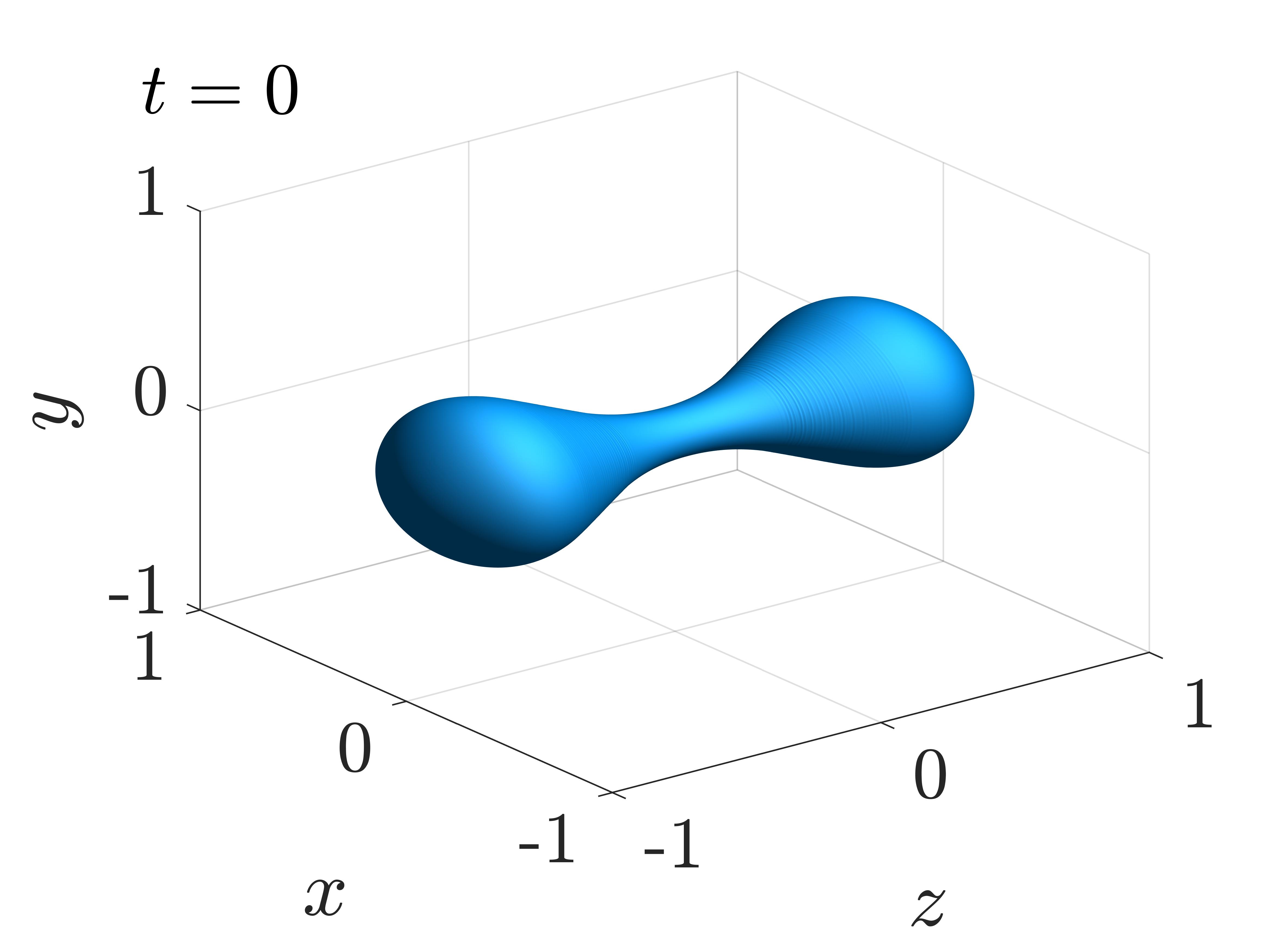}
	\includegraphics[width=0.3\linewidth]{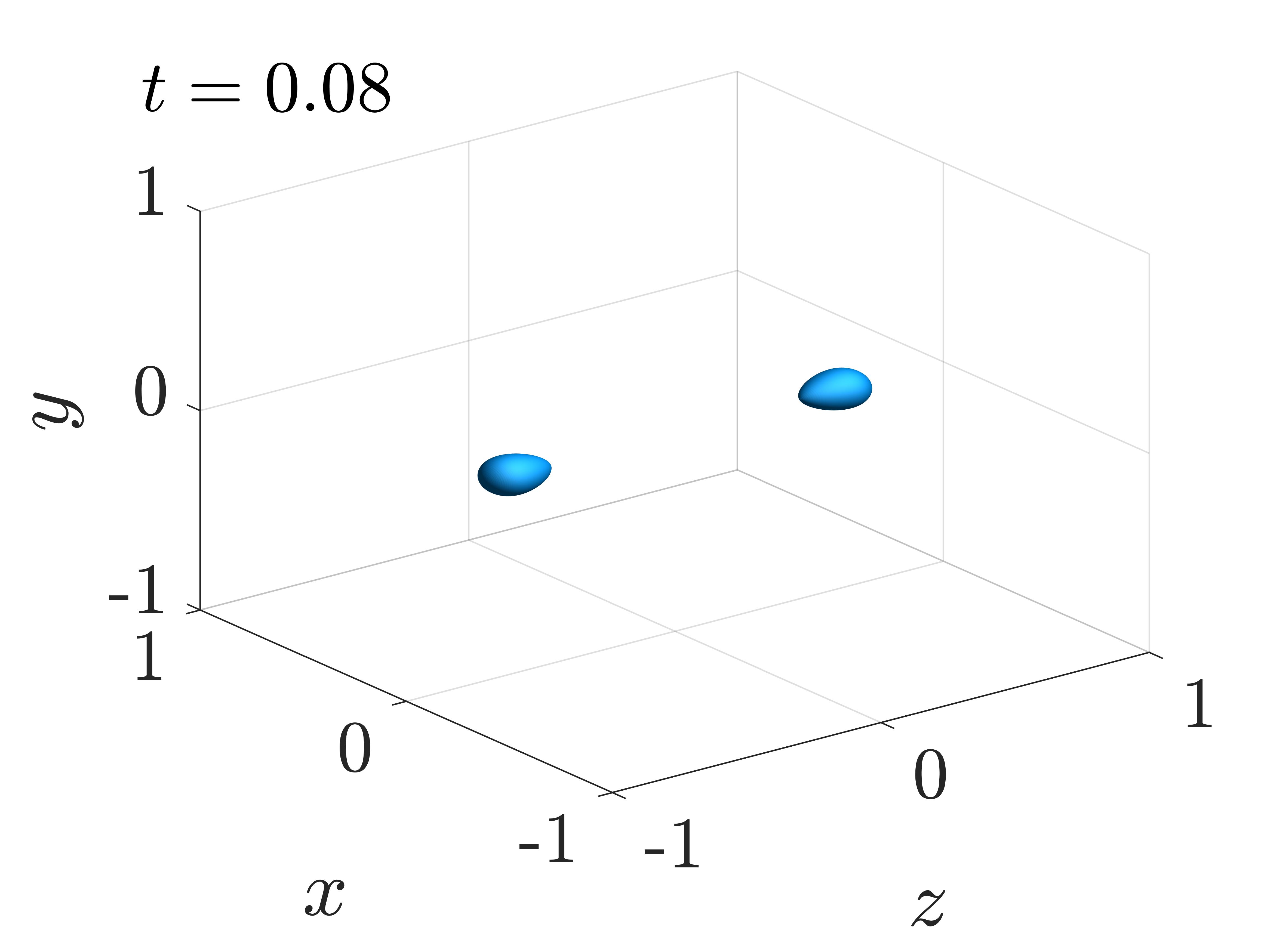}
	\includegraphics[width=0.3\linewidth]{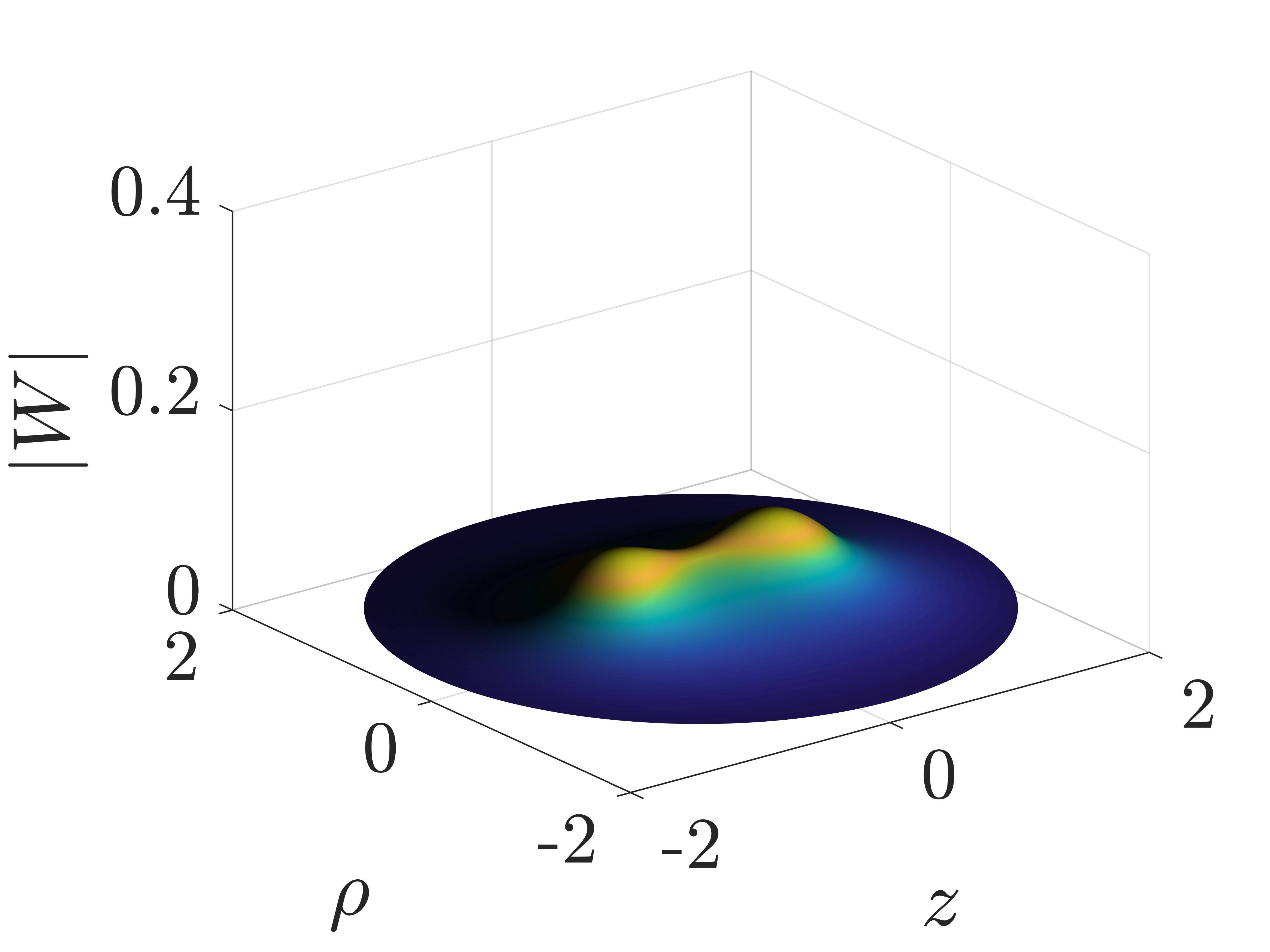}
	\caption{\label{fig:3DPlots} Numerical solution to \eqref{eq:zerostcondition}-\eqref{eq:Laplace} with initial conditions of the form \eqref{eq:example1}, \eqref{eq:examples2to4}, and \eqref{eq:SymmetricIC1a}-\eqref{eq:SymmetricIC1c}, and the corresponding numerical solution to \eqref{eq:Baiocchi1}-\eqref{eq:Baiocchi3}. Numerical solutions to \eqref{eq:zerostcondition}-\eqref{eq:Laplace} are computed using the level set based method described in \Cref{sec:NumericalScheme}, while the numerical solution to \eqref{eq:Baiocchi3} and \eqref{eq:Baiocchi5} is found using the procedure described in \Cref{sec:numericalzst}. Solutions are computed on the domain $0 \le \theta \le \pi$ and $0 \le r \le2$ using $628 \times 400$ equally spaced nodes. }
\end{figure}

In \Cref{fig:3DPlots}, we illustrate some numerical results by choosing parameter values from these three initial conditions.   For the prolate spheroid (\ref{eq:example1}) we provide results for $r_0=0.8$, noting that this initial condition is obviously convex.  For the peanut shaped surface (\ref{eq:examples2to4}), we choose $r_0=0.5$, which is not convex but is instead mean convex.  Finally, for the dumbbell shape (\ref{eq:SymmetricIC1a})-(\ref{eq:SymmetricIC1c}), we choose $r_0=4.75$, which again corresponds to a nonconvex shape which is still mean convex, but this time with a particularly thin neck region. In all of these case, we show in \Cref{fig:3DPlots} the initial shape, the numerical solution to \eqref{eq:zerostcondition}-\eqref{eq:Laplace} shortly before the extinction time, and the corresponding solution to the Baiocchi transform problem \eqref{eq:Baiocchi3} and \eqref{eq:Baiocchi5}.

For both of the first two examples in \Cref{fig:3DPlots}, namely \eqref{eq:example1} with $r_0 = 0.8$ and \eqref{eq:examples2to4} with $r_0 = 0.5$, the solution to \eqref{eq:zerostcondition}-\eqref{eq:Laplace} contracts to a single point at extinction.  By observing the third column of \Cref{fig:3DPlots}, we see this is consistent with the solution of \eqref{eq:Baiocchi3} and \eqref{eq:Baiocchi5}, which shows $|W|$ having one local maximum at the origin, predicting one point at extinction.  This comparison highlights that convex shapes and some nonconvex shapes will contract to a single point.  The extinction time predicted by the Baiocchi transform is computed by evaluating $|W|$ at ${\bf x}_\mathrm{e}$ (which, for this problem is the origin) giving the values $t_\mathrm{e} = 0.370$ and $t_\mathrm{e} = 0.233$ for \eqref{eq:example1} with $r_0 = 0.8$ and \eqref{eq:examples2to4} with $r_0 = 0.5$, respectively. Comparing this to the extinction times computed from the numerical solution to \eqref{eq:zerostcondition}-\eqref{eq:Laplace}, we find there is less than $0.1\%$ relative difference, suggesting excellent agreement.

\begin{figure}
	\centering
	\includegraphics[width=0.5\linewidth]{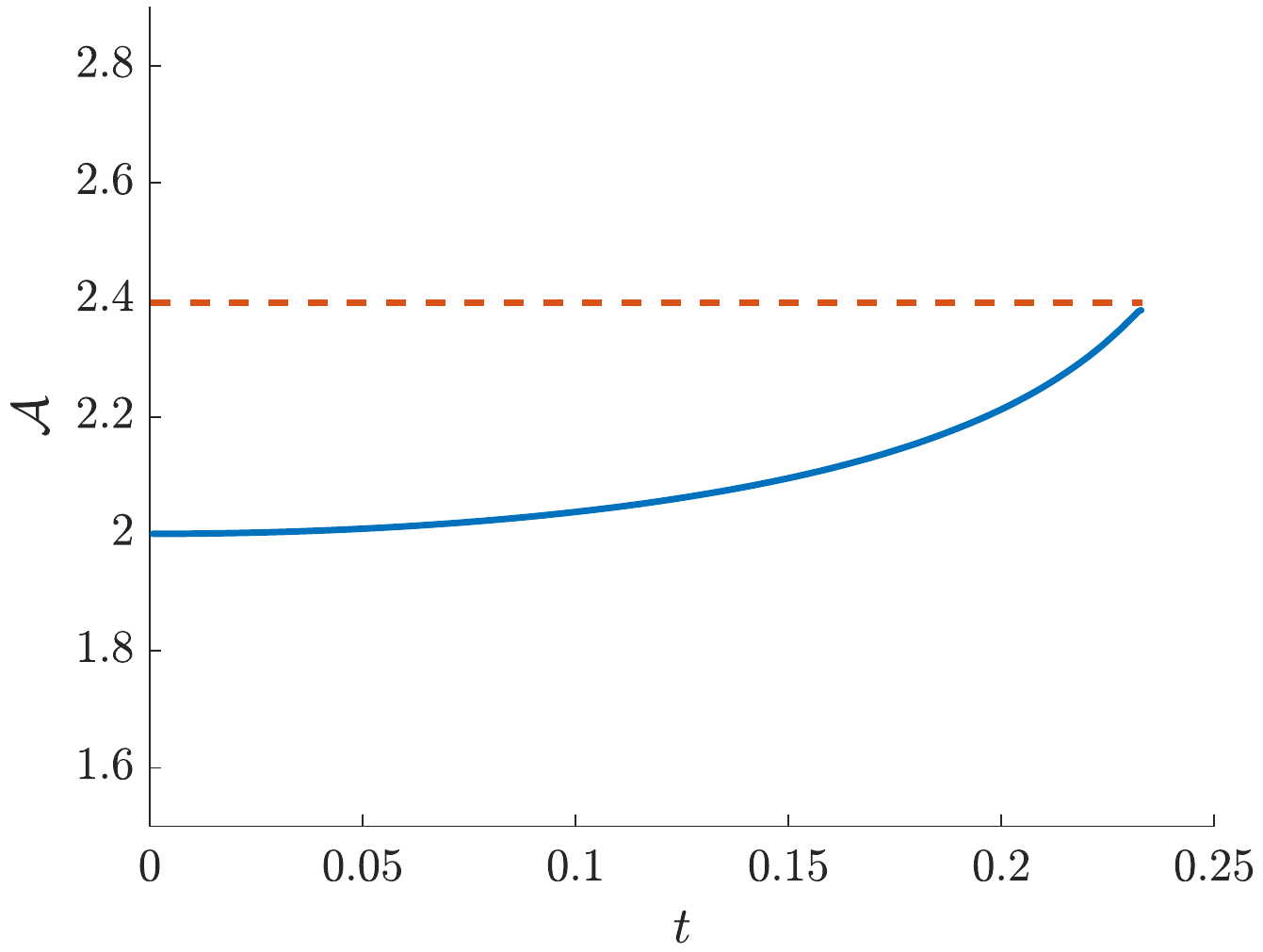}
	\caption{The evolution of the aspect ratio for the example initial condition (\ref{eq:examples2to4}) with $r_0=0.5$ is presented as a solid (blue) curve.  The (red) dashed curve is the predicted aspect ratio at extinction, given by (\ref{eq:aspectratio}). The agreement is quite good.}
	\label{fig:aspectratioexample}
\end{figure}

The equation (\ref{eq:examples2to4}) with $r_0=0.5$ provides a good test for the prediction (\ref{eq:aspectratio}).  For this purpose we take the solution to the Baiocchi transform problem (\ref{eq:Baiocchi}), which in this case predicts that $q_0=1.100$ and $a=0.215$.  As such, our prediction for the aspect ratio at extinction is $\mathcal{A}=2.395$.  The time-dependent behaviour of the aspect ratio for our numerical solution to the full problem (using the level set method) is presented in \Cref{fig:aspectratioexample}. This figure demonstrates how well these two results agree with other.

For initial condition \eqref{eq:SymmetricIC1a}-\eqref{eq:SymmetricIC1c} with $r_0=4.75$, \Cref{fig:3DPlots} shows different qualitative behaviour.  Here, we see that solutions to \eqref{eq:zerostcondition}-\eqref{eq:Laplace} will undergo pinch-off and ultimately the two satellite crystals will contract to separate points of extinction.  Again, this is consistent with the solution to \eqref{eq:Baiocchi3} and \eqref{eq:Baiocchi5} as the third column of \Cref{fig:3DPlots} indicates that $|W|$ has two local maxima.  By approximating the locations of these maxima and the values of $|W|$ at these points, we find the Baiocchi transforms predicts that the interface will contract to extinction points at $z_\mathrm{e}=\pm 0.577$ at time $t = 0.100$. Comparing these results with the extinction locations and times approximated from the numerical solution to \eqref{eq:zerostcondition}-\eqref{eq:Laplace}, we find a relative difference less than $0.2\%$.  This example shows, for symmetric initial conditions, how well the Baiocchi transform approach can be used to predict whether pinch-off will occur, as well as the extinction points and time.

In summary, these numerical results indicate that for a given initial interface, $\partial \Omega(0)$, each of the aspect ratio at extinction, the extinction time and location of the extinction point for an interface evolving according to \eqref{eq:zerostcondition}-\eqref{eq:Laplace} can be predicted from the solution to \eqref{eq:Baiocchi1}-\eqref{eq:Baiocchi3}.  Further, the indication is that this is true both for interfaces that contract to a single point of extinction, or undergo pinch-off and contract to multiple points of extinction, at least for symmetric initial conditions. Finally, these results illustrate the capacity of the level set based numerical scheme, presented in \Cref{sec:NumericalScheme}, to accurately describe the dynamics of the interface once a change in topology has occurred.

\subsubsection{Asymmetric initial condition}

The numerical solutions of \eqref{eq:zerostcondition}-\eqref{eq:Laplace} presented in \Cref{sec:SymmetricIC} indicate that when $\partial \Omega(t)$ is sufficiently non-convex then the interface will undergo a change in topology. As initial conditions considered in \Cref{sec:SymmetricIC} are symmetric along the major axis (about $\theta=\pi/2$), the two interfaces which form after pinch-off will have the same extinction time. We now investigate a class of asymmetric initial conditions that undergo pinch-off into two surfaces of differing volumes. We expect the smaller of the two volumes to contract to a point first, followed by the larger, thus giving two distinct extinction times.

\begin{figure}
	\centering
	\includegraphics[width=0.45\linewidth]{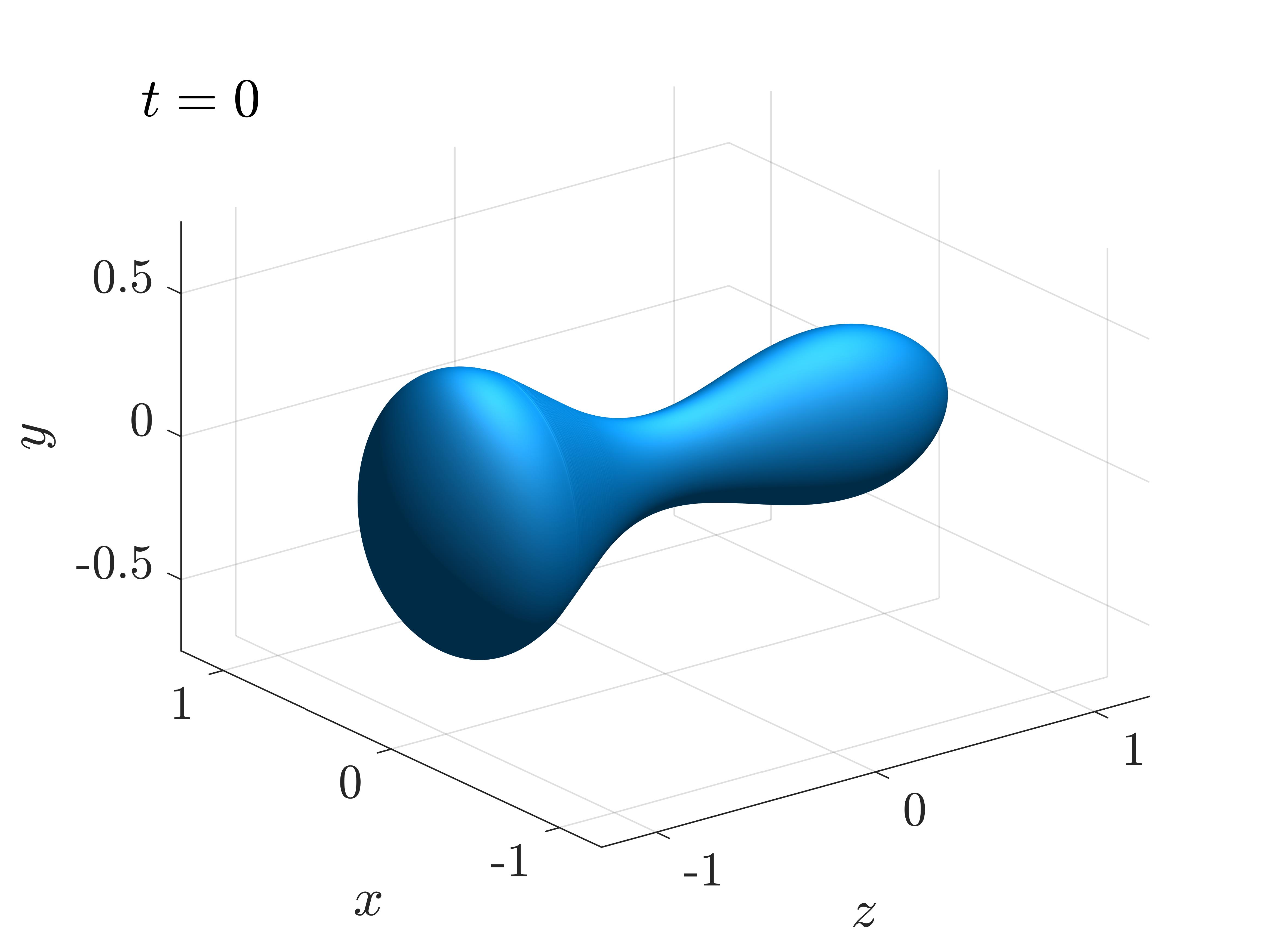}
	\includegraphics[width=0.45\linewidth]{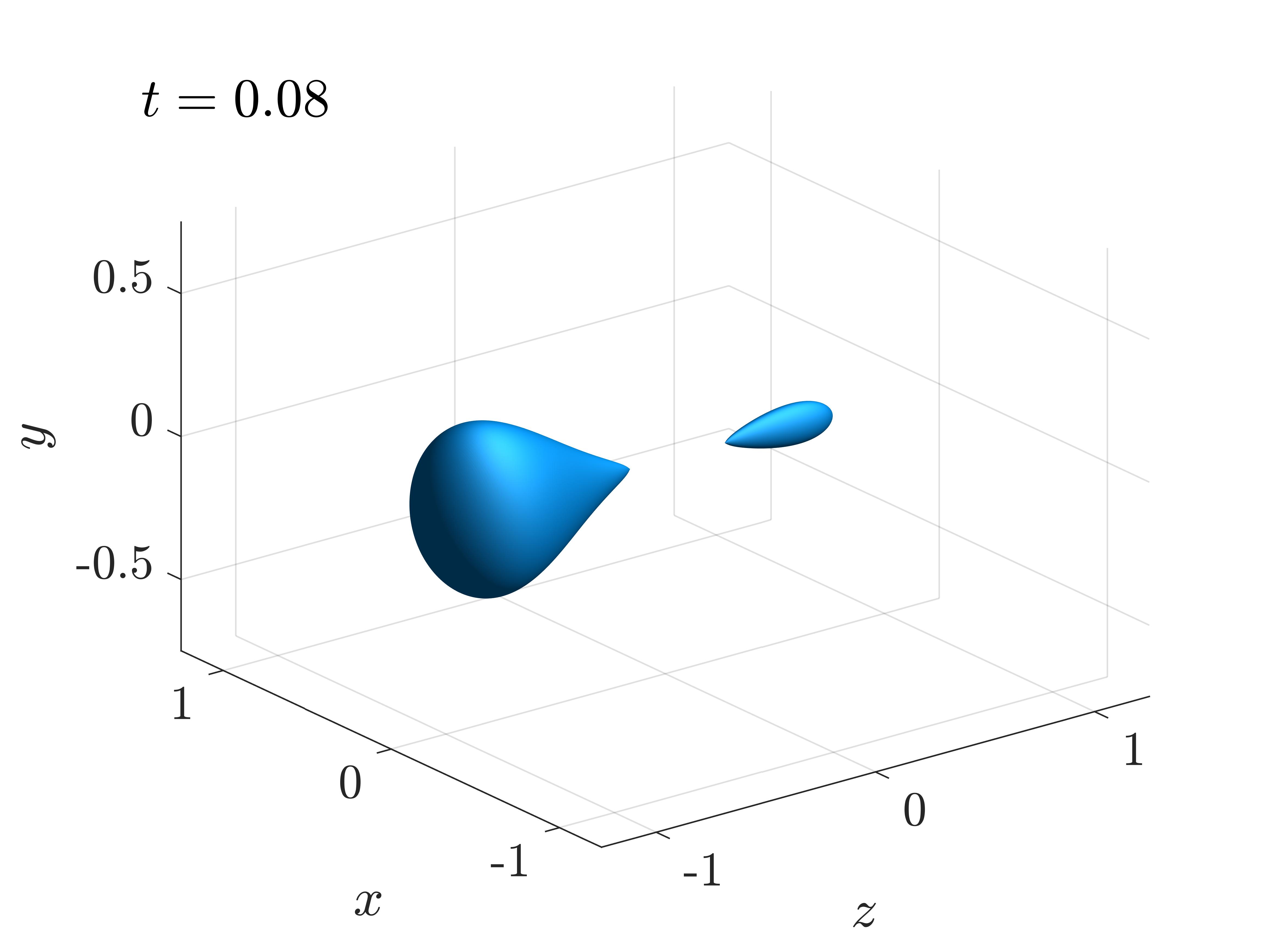}
	
	\includegraphics[width=0.45\linewidth]{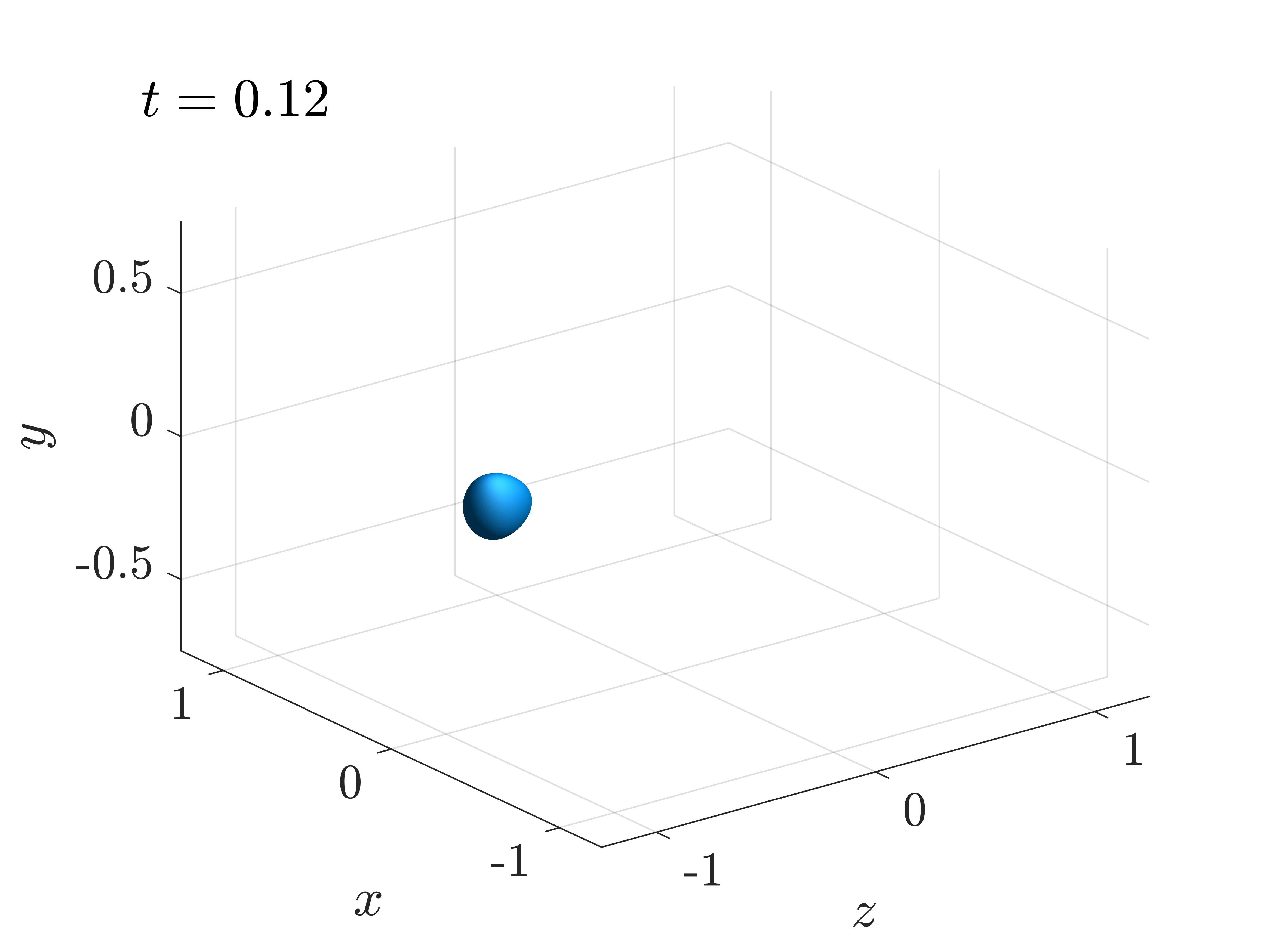}
	\includegraphics[width=0.45\linewidth]{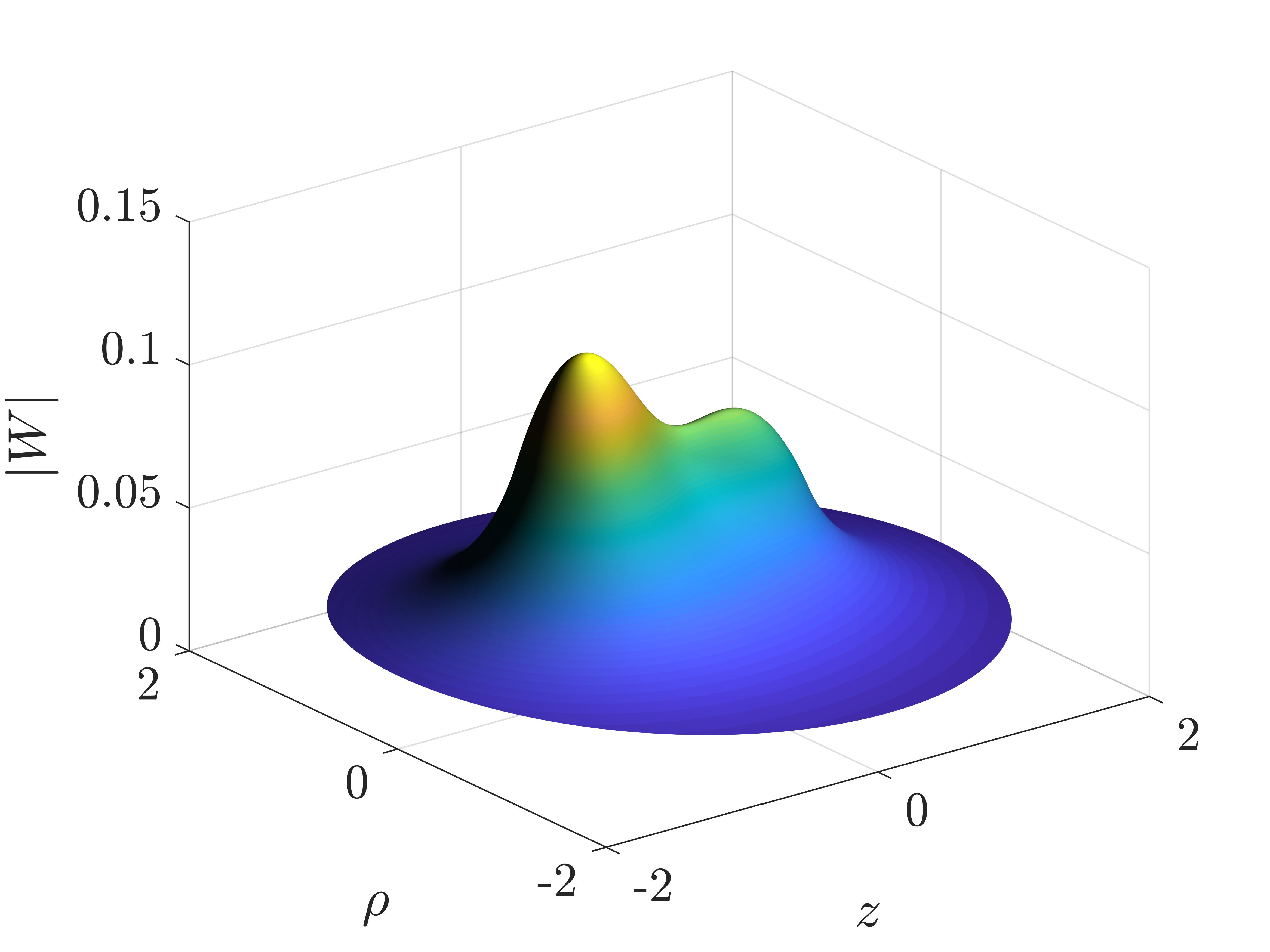}
	\caption{\label{fig:AsymmetricDumbbell} Time evolution of the numerical solution to \eqref{eq:zerostcondition}-\eqref{eq:Laplace} (computed using the level set based method described in \Cref{sec:NumericalScheme}), and corresponding numerical solution to \eqref{eq:Baiocchi1}-\eqref{eq:Baiocchi3} (found using the procedure described in \Cref{sec:numericalzst}). The initial condition is \eqref{eq:SymmetricIC1a}-\eqref{eq:SymmetricIC1c} with $r_0 = 5.1$. Solutions are computed on the domain $0 \le \theta \le \pi$ and $0 \le r \le2$ using $628 \times 400$ equally spaces nodes.}
\end{figure}

We again consider an initial condition of the form of \eqref{eq:SymmetricIC1a}-\eqref{eq:SymmetricIC1c}, but this time for $0 \le \theta \le \pi$.  In \Cref{fig:AsymmetricDumbbell}, we plot the time evolution of the numerical solution to \eqref{eq:zerostcondition}-\eqref{eq:Laplace} and the corresponding numerical solution to \eqref{eq:Baiocchi1}-\eqref{eq:Baiocchi3} for the representative case $r_0=5.1$. We observe that the full time-dependent solution to \eqref{eq:zerostcondition}-\eqref{eq:Laplace} undergoes a change in topology at approximately $t = 0.076$, with crystal domain $\Omega(t)$ pinching off into two.   The smaller of the two satellite crystals contracts to a point at $z_\mathrm{e} = 0.564$ when $t = 0.086$, followed by the remaining larger satellite crystal which contracts to a point at $z_\mathrm{e} = -0.773$ when $t = 0.127$.  The corresponding numerical solution to the Baiocchi transform problem \eqref{eq:Baiocchi5}, \Cref{fig:AsymmetricDumbbell} indicates that $|W|$ has two local maxima, located at $z_\mathrm{e}= 0.506$ and $z_\mathrm{e} = -0.767$, with $|W|$ equal to 0.092 and 0.127 at these points, respectively.  Thus we see that the predicted values of the extinction points and times agree well for the larger of the two satellite crystals (as it should) but not at all for the smaller crystal.  {red}{That our approach can only provide information about the extinction time and point for the largest satellite crystal} is a minor limitation to the Baiocchi transform framework.

\section{Effects of surface tension} \label{sec:SurfaceTension}

An inevitable consequence of melting a small crystal is that eventually the curvature will become large enough so that surface tension effects become important. For what follows, instead of \eqref{eq:zerostcondition} we use the dimensionless version of the Gibbs-Thomson law \eqref{eq:gibbsthoms}, which is
\begin{equation}
\mbox{on}\quad\partial\Omega: \qquad u=-\sigma \kappa,
\label{eq:Dynamic1}
\end{equation}
where $\sigma = \gamma u_m^{*} /\ell (u^*_\infty - u^*_m)$ is the dimensionless surface tension coefficient, and $\kappa$ is the dimensionless signed mean curvature.

\subsection{Linear stability analysis for near spherical crystal} \label{sec:LinearStability}

It proves useful to outline the linear stability analysis for interfaces evolving according to \eqref{eq:kinematicboundary}-\eqref{eq:Laplace} and \eqref{eq:Dynamic1} with a near-spherical initial condition.  In spherical polar coordinates $(r,\theta,\varphi)$, we represent the axially symmetric moving boundary $\partial\Omega$ by $r=s(\theta,t)$, so that our problem is
\begin{subequations}
	\begin{align}
	& \mbox{in} \quad r > s: && 0 = \frac{1}{r^2} \diffp{}{r} \left( r^2 \diffp{u}{r} \right) + \frac{1}{r^2 \sin \theta} \diffp{}{\theta} \left( \sin\theta \diffp{u}{\theta} \right), \label{eq:AxiallySymmetric1}\\
	& \mbox{on} \quad r = s: &&s_t = u_r - \frac{1}{s^2}u_\theta s_\theta, \label{eq:AxiallySymmetric2}\\
	& \mbox{on} \quad r = s: && u = \sigma \frac{ 3 s s_\theta^2 - \cot\theta s_\theta^3 - s^2 (s_{\theta\theta} + s_\theta \cot\theta  ) + 2s^3 }{s (s^2 + s_\theta^2)^{3/2}}, \label{eq:AxiallySymmetric3}\\
	& \mbox{as} \quad r \to \infty: && u \sim 1,  \label{eq:AxiallySymmetric4}
	\end{align}
\end{subequations}
We seek a perturbed spherical solution to \eqref{eq:AxiallySymmetric1}-\eqref{eq:AxiallySymmetric4} of the form
\begin{subequations}
	\begin{align}
	u(r, \theta, \varphi, t) &= u_0(r,t) + \varepsilon u_1(r, \theta, t) + \mathcal{O}(\varepsilon^2), \\
	s(\theta, t) &= s_0(t) + \varepsilon s_1(\theta, t) + \mathcal{O}(\varepsilon^2),
	\end{align}
\end{subequations}
where $\varepsilon \ll 1$.  The leading order solution is
\begin{equation}
u_0= 1 + \frac{2\sigma - s_0}{r}, \quad
s_0 = \frac{8 \sigma^2 \ln | (r_0 - 2\sigma)/(s_0 - 2\sigma)| + 2 t +  r_0(4 \sigma + r_0)}{4 \sigma + s_0}.
\end{equation}
where $s_0(0) = r_0$. For the $\mathcal{O}(\varepsilon)$ system,
\begin{subequations}
	\begin{align}
	& \mbox{in} \quad r > s_0: && 0 = \diffp{u_1}{r}\left( r^2\diffp{u_1}{r} \right) + \frac{1}{\sin \theta} \diffp{}{\theta} \left( \sin \theta \diffp{u_1}{\theta} \right),\\
	& \mbox{on} \quad r = s_0: && \diffp{s_1}{t} = \diffp{u_1}{r} + s_1\diffp[2]{u_0}{r},\\
	& \mbox{on} \quad r = s_0: && u_1 + s_1\diffp{u_0}{r} = - \sigma \frac{2s_1 + \cot \theta\partial_\theta s_1 + \partial^2_\theta s_1}{s_0^2}, \\
	& \mbox{as} \quad r \to \infty: && u_1\sim 0,
	\end{align}
\end{subequations}
the solutions are of the form
\begin{equation}
u_1(r, \theta, t) = \sum_{n=2}^{\infty} A_n r^{-n} P_n(\cos \theta), \qquad s_1(\theta, t) = \sum_{n=2}^{\infty} \gamma_n(t)P_n(\cos\theta)
\end{equation}
where $A_n$ is a sequence of unknown coefficients, $P_n$ is the $n$th Legendre polynomial, and $\gamma_n$ is the $n$th mode of perturbation to the sphere. We are able to eliminate $A_n$ to obtain
\begin{equation}
\frac{1}{\gamma_n}\frac{\mathrm{d}\gamma_n}{\mathrm{d}s_0} = \frac{(n-1)((n^2 + 3n + 4) \sigma + s_0)}{s_0(s_0 + 2\sigma)}.
\end{equation}
Since $(1/ \gamma_n)\mathrm{d}\gamma_n/\mathrm{d}s_0 \to 0$ in the limit that $s_0 \to 0$ for $n \ge 2$, we see that each mode of perturbation is stable, and a perturbed sphere will evolve to a sphere in the extinction limit, as expected.

The special case in which the initial condition is the prolate spheroid with major and minor axes $r_0 + \varepsilon$ and $r_0$, respectively, then
\begin{align} \label{eq:ProlateIC}
s(\theta, 0) &= \frac{r_0(r_0 + \varepsilon)}{\sqrt{(r_0\cos\theta)^2 + ((r_0+\varepsilon)\sin\theta)^2}}, \nonumber \\
&= r_0 + \varepsilon \left( \frac{1}{2} + \frac{2}{3}P_2 (\cos \theta)  \right) + \mathcal{O}(\varepsilon^2).
\end{align}
That is, $\gamma_2(0)=2/3$ and $\gamma_n(0)=0$ for $n\geq 3$.  This initial condition has an aspect ratio of $1 + \varepsilon / r_0 + \mathcal{O}(\varepsilon^2)$.  The exact solution for  $\gamma_2$ is
\begin{subequations}
	\begin{align}
	\gamma_2 &= \frac{2s_0^7}{3r_0^7} \left( \frac{r_0+2\sigma}{s_0+2\sigma} \right)^6,
	\end{align}
\end{subequations}
and the aspect ratio for this particular initial condition therefore becomes
\begin{equation} \label{eq:AR}
\mathcal{A}(s_0) = 1 + \varepsilon \frac{3\gamma_2}{2s_0} + \mathcal{O}(\varepsilon^2).
\end{equation}
Note that when $\sigma=0$, then $3 \gamma_2/ 2 s_0 = 1 / r_0$, resulting in the aspect ratio remaining constant, which is consistent with the known exact solution of \Cref{sec:exact}.  Otherwise, for $\sigma>0$, the aspect ratio decreases monotonically to unity, as shown later in \Cref{fig:linearstability1}.

\subsection{Long thin needle problem} \label{sec:needle}

We consider here the limit of a long thin melting dendrite. Suppose the axially-symmetric shape of the dendrite is given by $\rho = S(z,t)$ where $\rho^2 = x^2 + y^2$. Suppose also that $S_0(z) = S(z,0)$, $\rho_0(t) = S(0, t)$, $S(z_0(t),t) = 0$, where $\alpha = z_0(0) / \rho_0(0) \ll 1$ such that the initial aspect ratio, $\mathcal{A}(0) = 1/\alpha$, is large.

The inner region is for $r = \mathcal{O}(\rho_0(t))$. Here the melting is almost two-dimensional with $\partial u / \partial z \ll 1$ and $\partial S / \partial z \ll 1$ so that, to leading order,
\begin{subequations}
	\begin{align}
	& \mbox{in}\quad \rho > S(z,t): & \diffp[2]{u}{\rho} + \frac{1}{\rho} \diffp{u}{\rho} = 0, \label{eq:Needl1} \\
	& \mbox{on}\quad\rho=S(z,t): & u = -\frac{\sigma}{\rho}, \label{eq:Needl2}\\
	& \mbox{on}\quad\rho=S(z,t): & \diffp{S}{t} = -\frac{u}{\rho} \label{eq:Needl3}.
	\end{align}
\end{subequations}
The solution to \eqref{eq:Needl1}-\eqref{eq:Needl3} is
\begin{equation}
u = -S \diffp{S}{t} \ln(\rho/S),	\label{eq:innersolution}
\end{equation}
where the form for $S$ is determined by the missing far-field condition, which is found by considering the outer region.

In this outer region, which is for $r = \mathcal{O}(z_0(t))$, the dendrite appears as a slit. We scale $\tilde{\rho} = \rho / (\alpha \rho_0(t))$, $\tilde{t} / \ln \alpha$ and rewrite the inner solution \eqref{eq:innersolution} to be
\begin{equation} \label{eq:innersolution2}
u = -S \frac{\partial S}{\partial \tilde{t}} - \frac{\sigma}{S} - S\frac{\partial S}{\partial \tilde{t}} \frac{\ln(\rho_0 \tilde{\rho} / S)}{\ln \alpha}.
\end{equation}
The leading order solution in the outer region is $u=1$, thus, after matching with the leading order term in \eqref{eq:innersolution2} as $\alpha \to \infty$, we find
\begin{equation} \label{eq:innersolution3}
\frac{t}{\ln \alpha} = -\frac{1}{2} (S^2 - S_0^2) + \sigma (S - S_0) - \sigma \ln \left( \frac{S + \sigma}{S_0 + \sigma} \right).
\end{equation}
For the zero surface tension case $\sigma = 0$, we can solve \eqref{eq:innersolution3} explicitly to give
\begin{equation}
S(z,t) = \left( S_0^2 - \frac{2 t}{\ln \alpha} \right)^{1/2},
\end{equation}
again providing square root time dependence.

Of particular interest is the special case in which the initial dendrite is the prolate spheroid \eqref{eq:prolateinit}.
Here $\rho_0 = \alpha$ and $z_0(0) = 1$, so initially the dendrite has the aspect ratio $\mathcal{A}(0) = 1/\alpha$. From \eqref{eq:innersolution3} we find the interface is given implicitly by
\begin{equation}
1 - \frac{2 t}{\ln \alpha} = S^2 + \frac{z^2}{\alpha^2} + 2 \sigma \left[  \left( 1 - \frac{z^2}{\alpha^2} \right)^{1/2} - s + \ln \left( \frac{S+\sigma}{(1-z^2/\alpha^2)^{1/2} + \sigma} \right) \right].
\end{equation}
Note that the small parameter in this limit is $1/\ln \alpha$, which suggests the analysis here is valid only for extremely large aspect ratios.

\subsection{Numerical results for canonical problem} \label{sec:canonical}

For the melting prolate spheroidal crystal considered in \Cref{sec:baiocchi}, whose surface is \eqref{eq:prolate}, we find the mean curvature is largest near the tip, given by
\begin{equation}
\kappa = \frac{z_0(t)}{\rho_0(t)^2} = \frac{t_\mathrm{e}^{1/2}z_0(0)}{(t_\mathrm{e}-t)^{1/2}}.
\end{equation}
Thus the right hand side of \eqref{eq:Needl2} becomes $\mathcal{O}(1)$ when $t_\mathrm{e}-t = \mathcal{O}(\sigma^2)$, suggesting we rescale according to
\begin{equation}
t_\mathrm{e} - t = \sigma^2 \hat{t}, \quad \vec{x} = \sigma \vec{\hat{x}}, \quad u = \hat{u},
\end{equation}
and treat the following problem
\begin{subequations}
	\begin{align}
	& \mbox{in}\quad \mathbb{R}^3 \backslash \hat{\Omega}(\hat t) :  & \hat\nabla^2 \hat{u} = 0, \label{eq:SurfaceTensionModel1} \\
	& \mbox{on}\quad \partial\hat\Omega: & \hat{u} = -\hat{\kappa}, \label{eq:SurfaceTensionModel2} \\
	& \mbox{on}\quad \partial\hat\Omega: & \hat{v}_n = -\frac{\partial\hat{u}}{\partial\hat{n}},  \label{eq:SurfaceTensionModel3} \\
	& \mbox{as}\quad \hat{r}\rightarrow\infty: & \hat{u} \to 1,  \label{eq:SurfaceTensionModel4}
	\end{align}
\end{subequations}
when $\hat{t} = \mathcal{O}(1)$, $|\vec{\hat{x}}| = \mathcal{O}(1)$, where hats denote scaled quantities. For the case in which the initial crystal, $\hat{\Omega}(0)$, is a prolate spheroidal in shape, this is a canonical problem for melting a solid. This one parameter in the problem is the initial aspect ratio.

\begin{figure}
	\centering
	\includegraphics[width=0.4\linewidth]{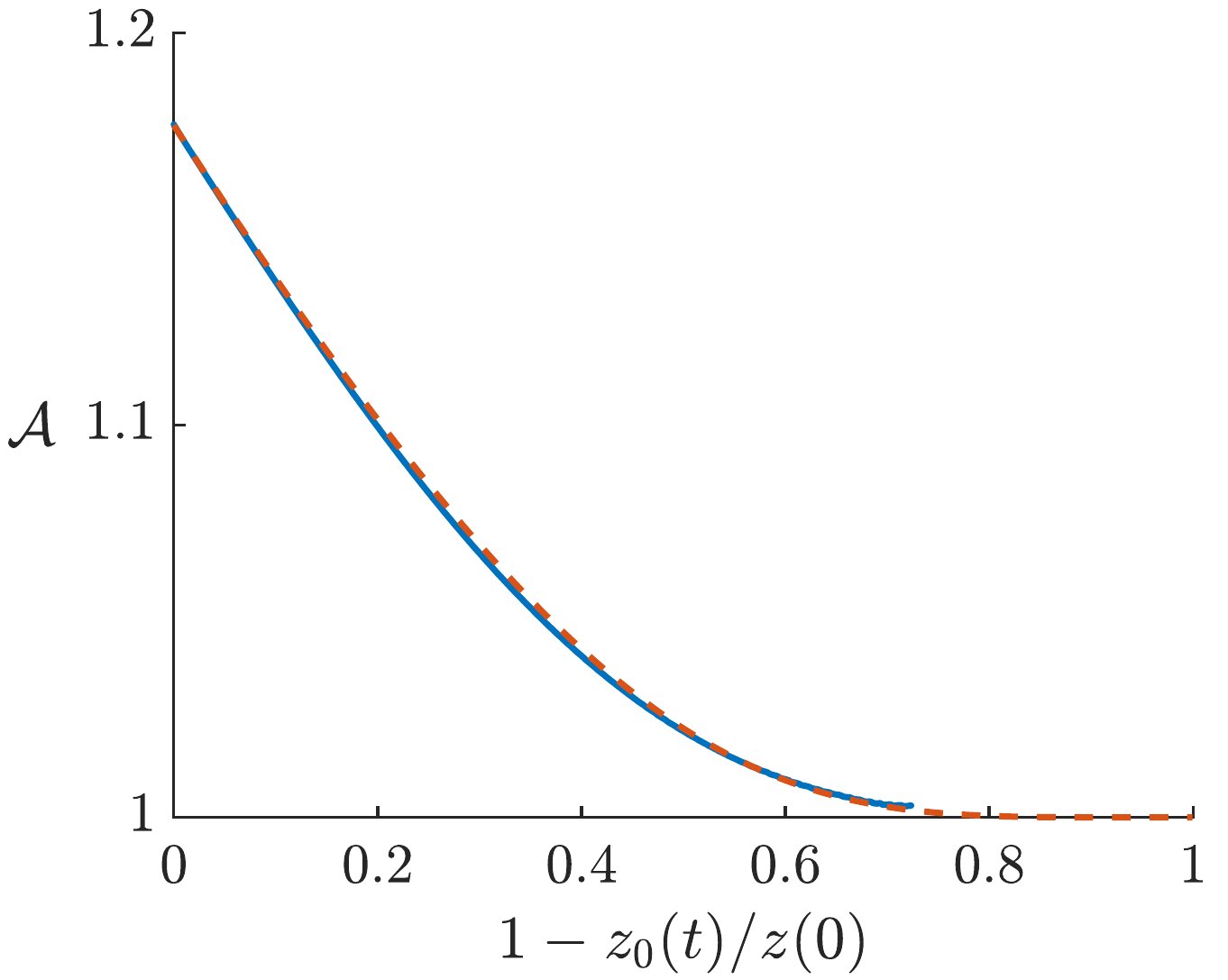}
	\caption{Comparison of the aspect ratio of the numerical solution to \eqref{eq:SurfaceTensionModel1}-\eqref{eq:SurfaceTensionModel4} (blue) with that predicted by linear stability analysis given by \eqref{eq:AR} (dashed red). Initial aspect ratio of the interface is $\mathcal{A}(0) = 20/17$. Numerical solution is computed on the domain $0 \le \theta \le \pi$ and $0 \le r \le 1.5$ with $314 \times 150$ equally spaced nodes.}
	\label{fig:linearstability1}
\end{figure}

\begin{figure}
	\centering
	\includegraphics[width=0.45\linewidth]{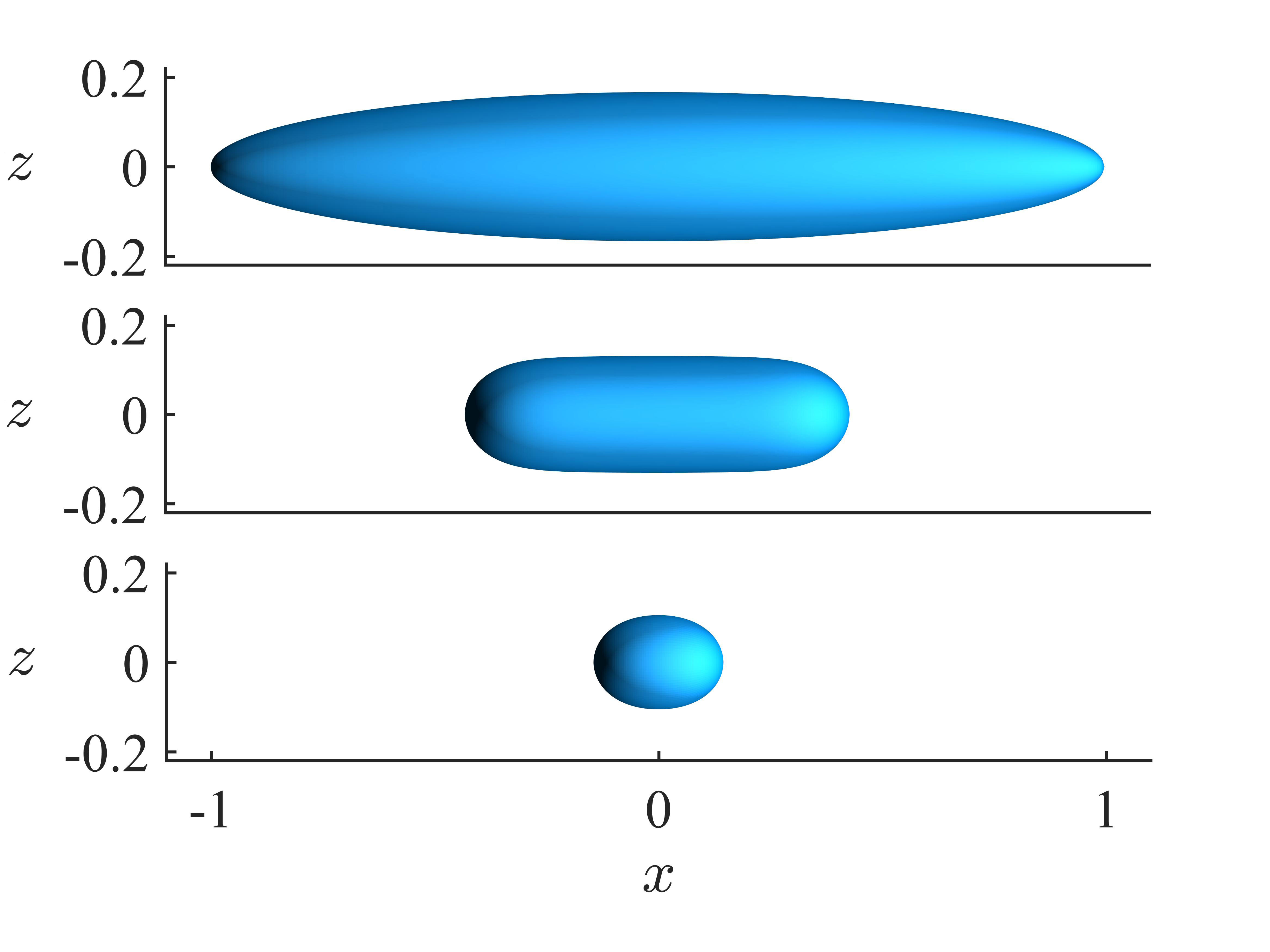}
	\includegraphics[width=0.45\linewidth]{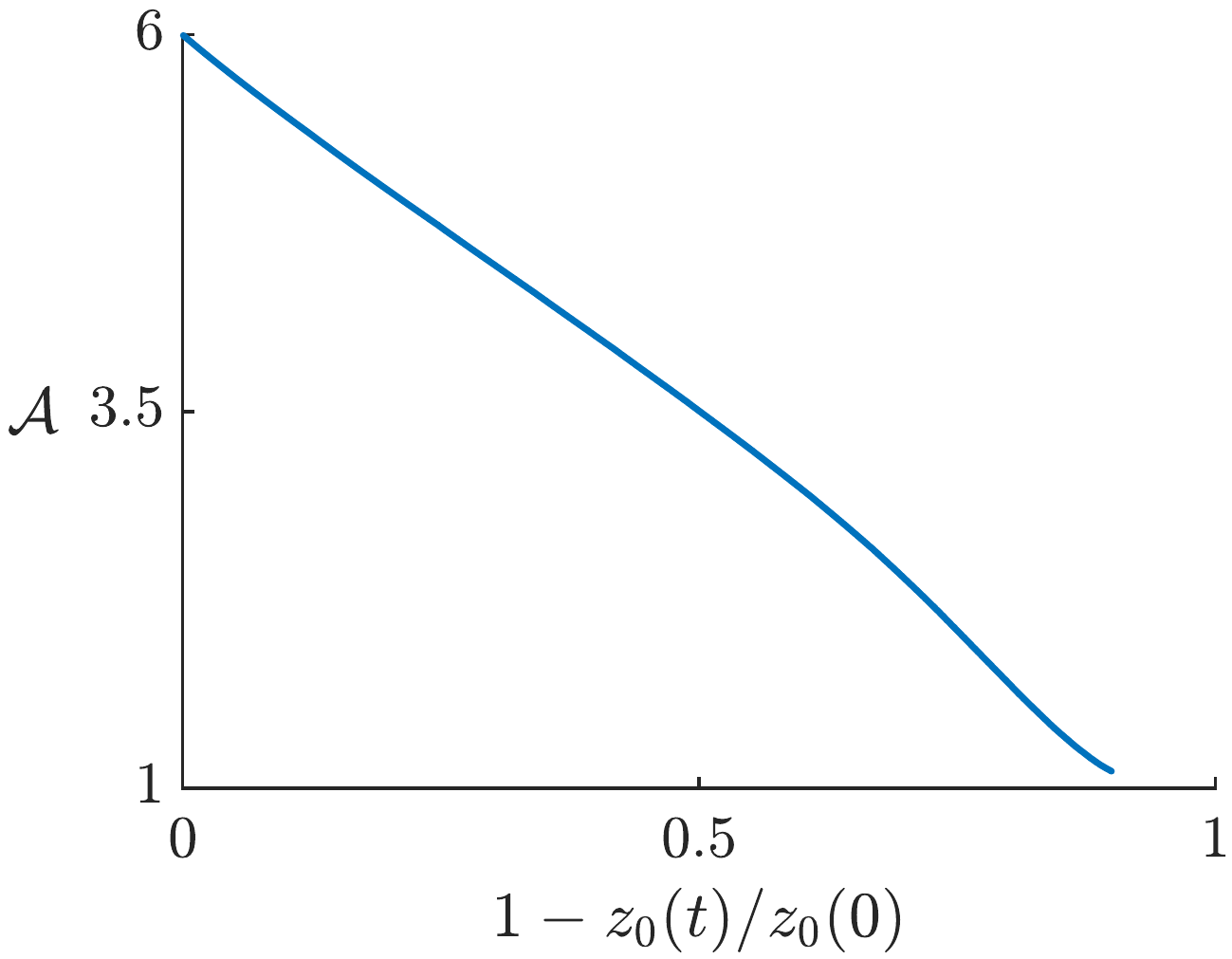}
	\caption{Left: Numerical solution to \eqref{eq:SurfaceTensionModel1}-\eqref{eq:SurfaceTensionModel4} at $t = 0$,  $0.0033$, and $0.0052$ computed using the scheme presented in \Cref{sec:NumericalScheme}. Initial condition is of the form \eqref{eq:example1} with $r_0 = 1/6$. Computations are performed on the domain $0 \le \theta \le \pi$ and $0 \le r \le 1.7$ with $624 \times 340$ equally spaced nodes.  Right: The corresponding aspect ratio as a function of time.}
	\label{fig:figure1aspectratio}
\end{figure}

Using the numerical scheme described in \Cref{sec:NumericalScheme}, we solve \eqref{eq:SurfaceTensionModel1}-\eqref{eq:SurfaceTensionModel4} for $\hat{u}$ and $\hat{\Omega}$. We first consider a near spherical prolate spheroid initial condition such that the initial aspect ratio is close to unity. \Cref{fig:linearstability1} compares the aspect ratio of the numerical solution to \eqref{eq:SurfaceTensionModel1}-\eqref{eq:SurfaceTensionModel4} with $\alpha = 0.85$ with the aspect ratio as predicted by linear stability analysis given by \eqref{eq:AR}. This figure shows excellent agreement between the numerical solution and linear stability analysis, confirming that the numerical scheme presented in \Cref{sec:NumericalScheme} is able to describe the behaviour of the interface as the aspect ratio decreases to unity. Further, we numerically solve \eqref{eq:SurfaceTensionModel1}-\eqref{eq:SurfaceTensionModel4} with $\alpha = 1/6$, and plot the time evolution of the solution and corresponding aspect ratio in \Cref{fig:figure1aspectratio}. As expected, this figure shows that the aspect ratio decays to unity in the limit that $t \to t_\mathrm{e}^-$.

\section{Kinetic undercooling} \label{sec:KineticUndercooling}

In this section, we very briefly consider the effects of extending the dynamic boundary condition \eqref{eq:Dynamic1} to include a kinetic undercooling-type term:
\begin{equation}
\mbox{on}\quad\partial\Omega: \qquad u=c v_n-\sigma \kappa,
\label{eq:Dynamic2}
\end{equation}
where $v_n$ is the normal velocity of $\partial \Omega$ and $c$ is the kinetic coefficient.  An argument for this extended boundary condition is that \eqref{eq:Dynamic1} can be derived under equilibrium conditions, while \eqref{eq:Dynamic2} is a corrected version that takes into account nonequilibrium kinetic effects~\cite{gurtin93,langer87}.  Physically, a nonzero kinetic coefficient $c>0$ penalises high interface speeds, which is important near extinction since our interface speed scales like $(t_\mathrm{e}-t)^{-1/2}$.  A wide variety of studies of Stefan problems have considered kinetic undercooling~\cite{back14a,back14b,davis01,ruiter17,evans00,font13,king05}.  The other important previous study is Dallaston \& McCue~\cite{dallaston13}, where the two-dimensional analogue of the quasi-steady problem (\ref{eq:Laplace}), (\ref{eq:Dynamic2}), (\ref{eq:kinematicboundary})-(\ref{eq:outerboundary}) is treated in some detail.

Following the linear stability analysis outlined in \Cref{sec:LinearStability} using \eqref{eq:Dynamic2} with $c>0$, we find the second mode of perturbation satisfies
\begin{equation}
\gamma_2 = \frac{2 s_0^2}{3 r_0^2} \left( \frac{3c+s_0}{2c+r_0} \right) ^{\frac{3c - 10\sigma}{3c - 2\sigma}} \left( \frac{r_0 +2 \sigma}{ s_0 + 2\sigma} \right)^{\frac{6(c-2\sigma)}{3c-2\sigma}},
\end{equation}
from which we see that
\begin{equation}
\lim\limits_{s_0\to 0^+} \frac{\gamma_2}{s_0} = 0,
\end{equation}
suggesting that an initially prolate spheroidal crystal will tend to a sphere in the extinction limit.  This conclusion is that same as before in \Cref{sec:LinearStability} when $c=0$.  On the other hand, a significant difference in qualitative behaviour is that the aspect ratio with $c>0$ may first increase and then decrease (to unity), which is a feature not observed when $c=0$.  The turning point can be calculated via
\begin{equation} \label{eq:TurningPoint}
\diff{}{s_0} \left(  \frac{\gamma_2}{s_0} \right) = 0 \quad \Rightarrow \quad s_0 = \frac{2\sigma c}{c - 4\sigma}.
\end{equation}
Given $s_0$ is defined on the domain $0 \le s_0 \le r_0$, the aspect ratio will monotonically decrease to unity if
\begin{equation}
\frac{2\sigma c}{c - 4\sigma}<0, \quad \textnormal{or} \quad r_0 > \frac{2\sigma c}{c - 4\sigma};
\end{equation}
otherwise, the aspect ratio will be non-monotone.

\begin{figure}
	\centering
	\includegraphics[width=0.45\linewidth]{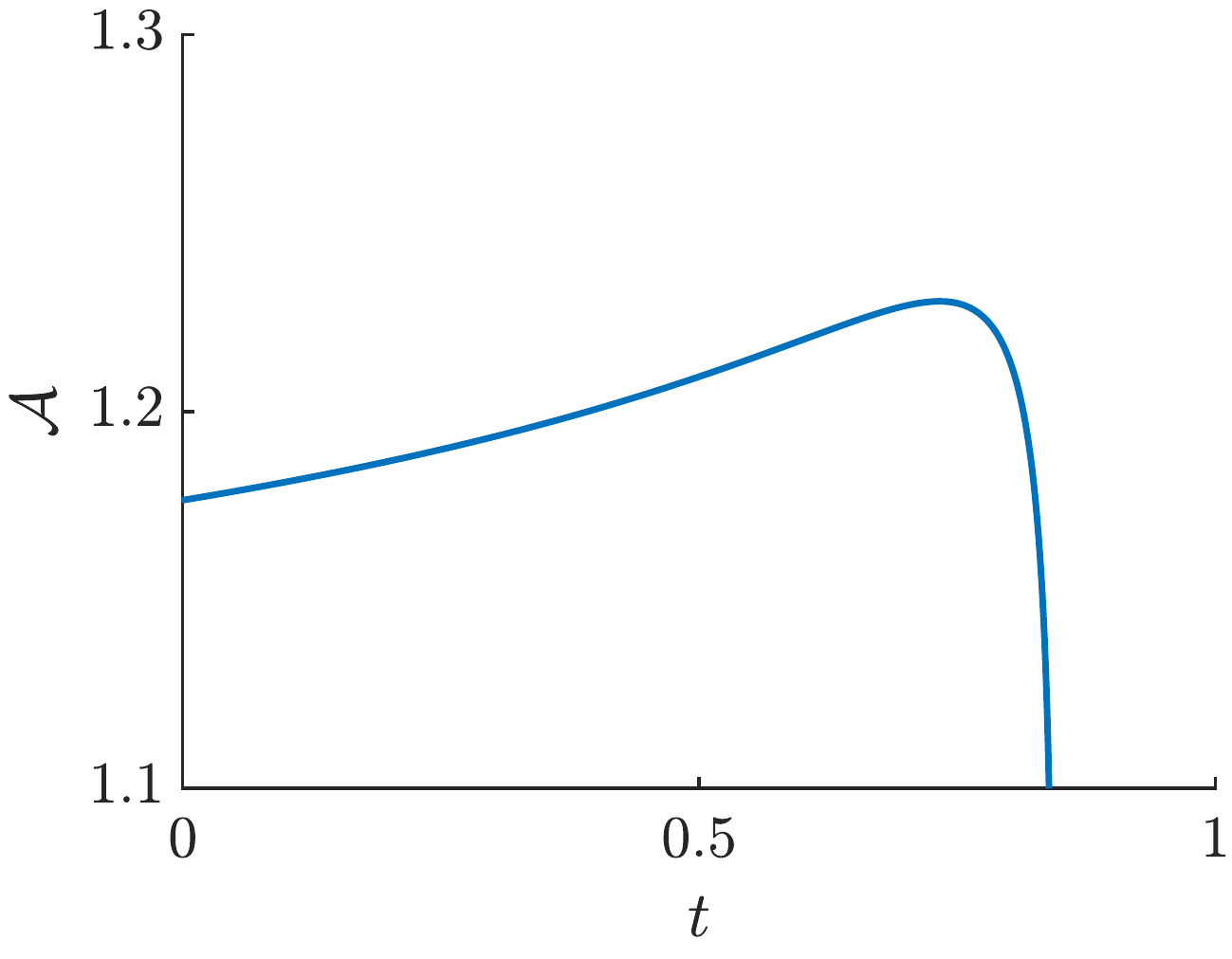}	
	\includegraphics[width=0.45\linewidth]{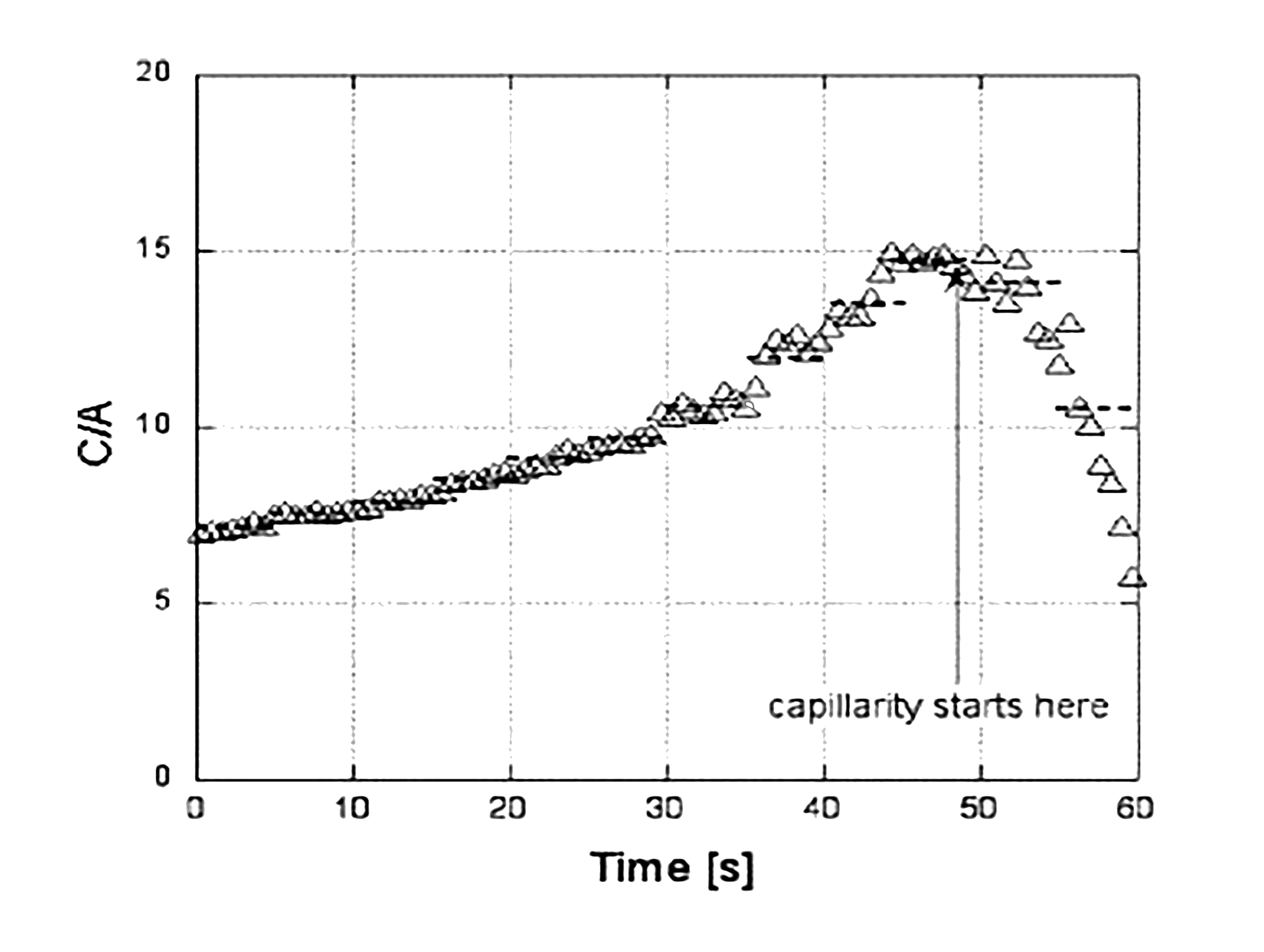}
	\caption{Left: The aspect ratio of a near spherical prolate spheroid as predicted by linear stability analysis from \eqref{eq:AR} with $\sigma = 0.075$ and $c = 1$. Right: The aspect ratio of a melting PVA crystal \cite{glicksman2}, reproduced with permission from Springer Nature.}
	\label{fig:AspectRatioComparison}
\end{figure}

Our work is motivated in part by a series of experiments performed as part of the IDGE \cite{glicksman0,glicksman2,glicksman}. In these experiments, it was observed that the aspect ratio of melting crystals increased for a period of time before decreasing to unity at extinction. In the context of the results presented in this section, \Cref{fig:AspectRatioComparison} illustrates the aspect ratio of a (near-spherical) prolate spheroid predicted by linear stability analysis and the aspect ratio of the melting PVA crystals \cite{glicksman2}. This figure shows that when both the effects of surface tension and kinetic undercooling are considered, the solution to \eqref{eq:kinematicboundary}-\eqref{eq:Laplace} and \eqref{eq:Dynamic2} is qualitatively similar to the experimental results (while of course the scale is different).

\section{Discussion} \label{sec:discussion}

In this paper, we have studied a quasi-steady one phase Stefan problem for melting an axially symmetric crystal.  In \Cref{sec:extinction} we treat a zero-surface-tension model and use analytical tools to show that axially symmetric crystals will tend to prolate spheroids in the limit that they melt completely, namely $t \to t_\mathrm{e}^-$, with an aspect ratio that depends on the initial condition.  The point to which the crystals ultimately shrink, together with the melting time, is predicted by this analysis and confirmed using a novel numerical scheme based on the level set method (presented in \Cref{sec:NumericalScheme}).  An advantage of this scheme is that we are also able to present numerical results for crystals that undergo pinch-off and contract to multiple points of extinction.

We consider the effects of surface tension by the Gibbs-Thomson law (\ref{eq:gibbsthoms}) in \Cref{sec:SurfaceTension}.  By performing linear stability analysis on the spherical solution, we show that surface tension acts to smooth out perturbations to the interface, suggesting it becomes spherical in the extinction limit.  A numerical study of canonical problem confirms this prediction.  These results are as expected and also indicated by the experimental results summarised by Glicksman and co-workers \cite{glicksman0,glicksman2,glicksman}.  However, the one feature of the IDGE not described by the model with surface tension is the non-monotonic behaviour of the aspect ratio, where the aspect ratio first increases as the crystal becomes very long and thin, and then very quickly decreases to unity as surface tension ultimately acts to produce a perfect sphere in the extinction limit.  In order to mimic this non-monotonic behaviour, we have included the effects of kinetic undercooling in the model in \Cref{sec:KineticUndercooling}, which shows that the competition between kinetic effects and surface tension does indeed produce the qualitative behaviour observed.

A key assumption in our paper is that the Stefan number in (\ref{eq:heat_u}) is taken to be large, namely $\beta\gg 1$, so that (\ref{eq:heat_u}) reduces to (\ref{eq:Laplace}) and our moving boundary is therefore quasi-steady.  There are two issues related to this assumption that we wish to mention.  First, our problem for melting a crystal is the same as that for a bubble contracting in a porous medium where the flow is governed by Darcy's law \cite{dibenedetto,howison1,mccue2}, although in that context the far-field (Dirichlet-type) boundary condition (\ref{eq:outerboundary}) should probably be replaced with a flux condition that dictates how quickly the bubble volume is decreasing (in two dimensions the equations describe bubble contraction in a Hele-Shaw cell \cite{entov,entov2,lee}).  For the case in which a bubble pinches off to produce two shrinking bubbles, the problem formulation would also need to consider two points of extraction that coincide with the eventual extinction points.  The second issue is that, strictly speaking, for the extremely late stages of melting, our quasi-steady model with (\ref{eq:Laplace}) is no longer applicable in the large Stefan number limit, and instead (\ref{eq:heat_u}) must be retained.  The mathematical details of such an exponentially short final-melting stage have been recorded in a number of previous studies \cite{andreucci01,herrero97,mccue1,mccue3,soward}.

\section*{Acknowledgments}
SWM and LCM acknowledge the support of the Australian Research Council Discovery Project DP140100933.  We are grateful to the anonymous referees for their helpful feedback.

\appendix
\section{Prolate spheroids with constant aspect ratio}\label{sec:appendix}

To solve the inner problem (\ref{eq:baiocchi11})-(\ref{eq:baiocchi13}) we employ prolate spheroidal coordinates $(\xi,\eta,\phi)$ defined by
\begin{subequations}
	\begin{align}
	X &= k \sinh \xi \sin\eta \cos\phi\\
	Y &= k \sinh \xi \sin\eta \sin\phi \\
	Z &= k \cosh \xi \cos\eta,
	\end{align}
\end{subequations}
where $\xi\geq 0$, $0\leq\eta\leq\pi$, $0\leq \phi < 2\pi$, and $k$ is a constant to be determined below.  The crystal boundary $\partial\Omega_0$ is described by $\xi=\xi_0$ or, equivalently,
\begin{equation}
\frac{X^2+Y^2}{\sinh^2\xi_0}+\frac{Z^2}{\cosh^2\xi_0}=k^2.
\end{equation}
Motivated by the relationship
\begin{equation}
a(X^2  +Y^2) + \left(\frac{1}{2}-2a\right)Z^2=  \frac{1}{2}k^2\left[\left(\frac{1}{2}-a\right)\cosh^2\xi
-a\right] +\frac{1}{2}k^2\left[\left(\frac{1}{2}-3a\right)\cosh^2\xi+a\right]\cos 2\eta,
\end{equation}
we look for a solution of the form
\begin{equation}
\Phi=f_1(q)+f_2(q)\cos 2\eta,
\end{equation}
where $q=\cosh\xi$ and $q_0=\cosh\xi_0$ and obtain a coupled system of two second order (Legendre-type) differential equations for $f_1$ and $f_2$.  These (and the constant $k$) are solved subject to the four conditions $f_1=f'_1=f_2=f'_2=0$ on $q=q_0$, and the far-field condition (\ref{eq:baiocchi13}) to give
\begin{alignat}{2}
f_1 &=& \frac{1}{2}k^2\left[\left(\frac{1}{2}-a\right)q^2-a\right]
-d + \frac{1}{8}k^2 q_0(q_0^2-1)\left[q-\frac{1}{2}(q^2-3)\ln\left(\frac{q+1}{q-1}\right)\right],
\\
f_2&=&
\frac{1}{2}k^2\left[\left(\frac{1}{2}-3a\right)q^2+a\right]
-d + \frac{1}{8}k^2 q_0(q_0^2-1)\left[3q-\frac{1}{2}(3q^2-1)\ln\left(\frac{q+1}{q-1}\right)\right],
\end{alignat}
where
\begin{equation}
k=q_0^{-1/3}(q_0^2-1)^{-1/3},
\end{equation}
and $d$ is given by (\ref{eq:d}).  The important relationship between $q_0$ and the special constant $a$ is given by (\ref{eq:a}).

\section{Numerical solution - A level set approach} \label{sec:NumericalScheme}

To find the numerical solution of \eqref{eq:zerostcondition}-\eqref{eq:Laplace}, we implement a level set based approach. The level set method (LSM), first proposed by Osher and Sethian \cite{Osher1988}, is a tool used to study a wide range of moving boundary problems. We refer the reader to Osher \& Fedkiw~\cite{Osher2003} and Sethian~\cite{Sethian1999} for comprehensive overviews of implementation strategies and applications. The LSM utilises an Eulerian approach by representing an $n$-dimensional interface, $\partial \Omega(t)$, as the zero level set of a $n+1$-dimensional surface, $\phi(\vec{x}, t)$, such that
\begin{equation}
\partial \Omega(t) = \left\lbrace  \vec{x} | \phi(\vec{x}, t) = 0  \right\rbrace .
\end{equation}
By representing the interface implicitly, the LSM can be used to describe complex behaviour such as the changes in topology observed in \Cref{fig:3DPlots}, while operating on a simple regular two-dimensional grid.

The evolution of the level set function $\phi$ is described by the level set equation
\begin{equation} \label{eq:LevelSetEqn}
\diffp{\phi}{t} + F|\nabla \phi| = 0\,,
\end{equation}
where $F$ is a continuous function defined on all of the computational domain, satisfying $F = V_n$ on $\vec{x} = \partial \Omega(t)$.  In the context of \eqref{eq:zerostcondition}-\eqref{eq:Laplace}, by noting that the outward normal of $\phi$ is $\vec{n} = \nabla \phi / |\nabla \phi|$, a suitable expression for $F$ on and outside the interface is
\begin{equation}  \label{eq:SpeedFunction}
F =  \frac{\nabla u \cdot \nabla \phi}{| \nabla \phi |} \qquad \vec{x} \in \mathbb{R}^3 \backslash \Omega(t).
\end{equation}

This leaves the matter of defining a suitable extension of $F$ to inside the interface.  Among several possibilities in the level set literature, we opt for a biharmonic extension as proposed by Moroney et al.~\cite{moroney}, and compute $F$ inside the interface to satisfy
\begin{equation} \label{eq:Biharmonic}
\nabla^4 F = 0 \qquad \vec{x} \in \Omega(t),
\end{equation}
together with the boundary conditions that $F$ and $\partial F/ \partial n$ are continuous across $\partial \Omega(t)$.  This method of extension shares the main property of the LSM itself, in not requiring the location of the interface to be calculated explicitly.  To solve \eqref{eq:Biharmonic}, we formulate the biharmonic stencil over the entire domain, which is then modified so that values of $F$ outside the interface, whose location is determined from the sign of $\phi$, are not overwritten. The resulting linear system is solved using LU decomposition. This extension is a variant of a two-dimensional thin plate spline interpolant defined on the level set grid.

\subsection{General algorithm}

The algorithm used to solve \eqref{eq:zerostcondition}-\eqref{eq:Laplace} numerically is outlined as follows:
\begin{itemize}
	\item[\textit{Step 1}] For a given initial condition $s(\theta,0)$, construct a level set function $\phi(r,\theta,0)$ such that $\phi<0$ inside the interface and $\phi>0$ outside the interface. This function is then converted to a signed distance function using the method of crossing times as described by Osher \& Fedkiw~\cite{Osher2003}.
	\item[\textit{Step 2}] Compute the temperature, $u$, on the domain $r \ge s(\theta,t)$ using the procedure described in \Cref{sec:ComputingTemp}.
	\item[\textit{Step 3}] Compute $F$ according to \eqref{eq:SpeedFunction}, where the derivatives are evaluated using central finite differences. $F$ is extended over the entire computational domain by solving \eqref{eq:Biharmonic} at nodes where $\phi<0$, with boundary data from step 3.
	\item[\textit{Step 4}] Update $\phi$ by advancing the level set equation given by \eqref{eq:LevelSetEqn}, where the time step is $\Delta t = 0.25 \times \Delta x / \max |F|$. We discretise the spatial derivatives in \eqref{eq:LevelSetEqn} using a ENO2 scheme for the spatial derivatives and integrate in time using second order Runge-Kutta where $\Delta t = 0.25 \times \Delta r / \max|F|$.
	\item[\textit{Step 5}] Reinitialise $\phi$ every 5 time-steps to a signed distance function by solving the reinitialisation equation
	\begin{equation} \label{eq:Reinit}
	\partial_\tau \phi + S(\phi)(|\nabla \phi | - 1) = 0,
	\end{equation}
	where
	\begin{equation}
	S(\phi) = \frac{\phi}{\sqrt{\phi^2 + \Delta r^2}}.
	\end{equation}
	We use 5 pseudo-timesteps with $\Delta \tau = 0.2 \Delta r$.
	\item[\textit{Step 6}] Repeat steps 2-5 until the desired simulation time is attained.
\end{itemize}

\subsection{Solving for temperature} \label{sec:ComputingTemp}

\begin{figure}
	\centering
	\includegraphics[width=0.45\linewidth]{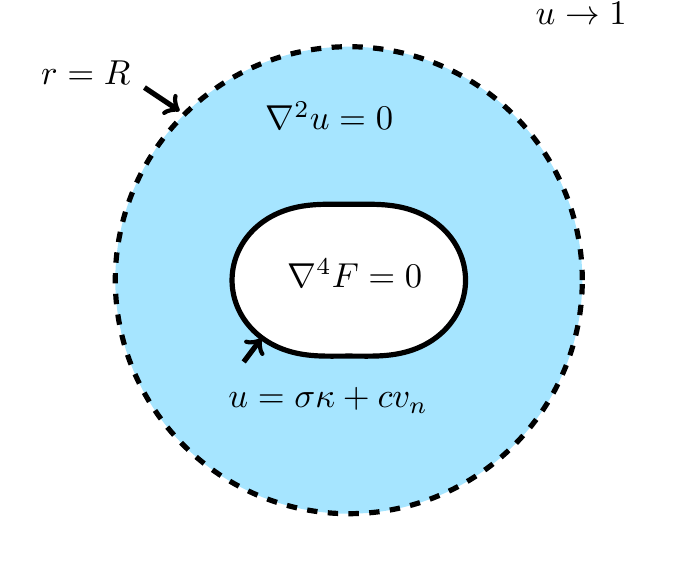}
	\caption{Schematic of how the speed function, $F$, is computed for each time step. Blue region denotes where temperature, $u$, is solved for using finite differences. This finite difference stencil must be adjusted to incorporate the dynamic boundary condition (\Cref{sec:FieldEqn}). To incorporate the far-field boundary condition, we impose an artificial boundary at $r = R$ and implement a Dirichlet to Neumann mapping (\Cref{sec:Farfield}). $F$ is computed outside the interface using \eqref{eq:SpeedFunction}, and is extended to be defined over the entire computational domain by solving the biharmonic equation.}
	\label{fig:SolveF}
\end{figure}

Evaluating the speed function $F$ in the level set equation \eqref{eq:LevelSetEqn} requires first calculating the temperature $u$.  This is achieved by using a modified finite difference stencil for Laplace's equation in the region outside the interface. For nodes away from the interface, a standard 5-point stencil is used such that the discrete equation is
\begin{equation}\label{eq:FiniteDifference}
\begin{split}
0 &= \frac{u_{i-1,j} - 2u_{i,j} + u_{i+1,j}}{\Delta r^2} + \frac{2}{r_{i,j}} \frac{u_{i+1,j}-u_{i-1,j}}{2\Delta r}
\\
&
+ \frac{1}{r_{i,j}^2} \frac{u_{i,j-1} -2u_{i,j} +u_{i,j+1}}{\Delta \theta^2}
+ \frac{\cot \theta}{r_{i,j}^2} \frac{u_{i,j+1}-u_{i,j-1}}{2\Delta \theta}.
\end{split}
\end{equation}
For the singularity at $\theta = 0$, noting that $\partial u / \partial \theta = 0$ and using L'Hoptial's rule then
\begin{align}
\lim\limits_{\theta \to 0^{+}} \cot \theta  \frac{\partial u}{\partial \theta} = \frac{\partial^2 u}{\partial \theta^2}.
\end{align}
The same procedure is applied at $\theta = \pi$. Difficulties arise when attempting to incorporate the dynamic condition \eqref{eq:zerostcondition} on the interface and the far-field boundary condition \eqref{eq:outerboundary}. We detail the methodology used to overcome each of these difficulties in \Cref{sec:FieldEqn,sec:Farfield}, respectively. A schematic of the problem is given in \Cref{fig:SolveF}, which illustrates the different equations to be solved in each part of the computational domain.

We note that since the governing equation for temperature satisfies Laplace's equation, an alternative approach for computing the temperature $u$ is the boundary integral method, which can be coupled with the level set method to solve problems where changes in topology occur \cite{Garzon2011}. However, an advantage of using a finite difference stencil is that it can easily be adapted to problems where the boundary integral method is not applicable. For example, we have used a similar method to the one presented in this section to study non-standard Hele-Shaw flow where pressure is not harmonic and for which the boundary integral method is much less suitable \cite{Morrow2019}.

\subsubsection{Incorporating the dynamic boundary condition} \label{sec:FieldEqn}

Special consideration must be taken when solving for nodes adjacent to the interface as we can no longer use the second order central differencing scheme \eqref{eq:FiniteDifference}. Instead we follow the work of Chen et al.~\cite{chen} and approximate the spatial derivatives by fitting a quadratic polynomial from values on and near the interface and differentiating this polynomial twice. Supposing the interface is located between two nodes $(i-1,j)$ and $(i,j)$, the quadratic is fitted using the three points $(r_b, u_b)$, $(r_{i,j},u_{i,j})$, and $(r_{i+1,j},u_{i+1,j})$. Here $r_b$ denotes the location of the interface and $u_b$ is the temperature at the interface. The value of $r_b$ is found by noting that $\phi$ is a signed distance function and so the distance between $r_b$ and $r_{i,j}$, denoted $h$, can be calculated by
\begin{equation}
h = \Delta r \left| \frac{\phi_{i,j}}{\phi_{i,j} - \phi_{i-1,j}} \right|.
\end{equation}
Thus \eqref{eq:FiniteDifference} becomes
\begin{equation}
\begin{split}
\diffp[2]{u}{r} + \frac{2}{r} \diffp{u}{r} \to \left( \frac{2}{h(h+\Delta r)} - \frac{2}{r_{i,j}} \frac{\Delta r}{h (\Delta r + h)} \right) u_b+\left( \frac{2}{r_{i,j}} \frac{\Delta r - h}{h \Delta r} -\frac{2}{h \Delta r}\right)u_{i,j}  \\ +  \left( \frac{2}{\Delta r (h + \Delta r)} + \frac{2}{r_{i,j}}\frac{h}{\Delta r (h + \Delta r)} \right) u_{i+1,j}.
\end{split}
\end{equation}
The same procedure is applied if the interface is between $r_i$ and $r_{i+1}$, or in the azimuthal direction.

The value of $u_b$ is determined by the dynamic condition \eqref{eq:Dynamic2}, where in the case of surface tension the mean curvature term
\[ \kappa = \nabla \cdot \left(  \frac{\nabla \phi}{|\nabla \phi|} \right)  \]
is approximated using central finite differences, while the normal velocity from the previous time step is used for the kinetic undercooling term.

\subsubsection{The far-field condition} \label{sec:Farfield}

Special consideration must also be given when considering the boundary condition at $r \to \infty$. One method for simulating this far-field condition is to make the computational domain much larger than the radius of the interface and then impose $u = 1$ on the outer boundary.  However, this is computationally expensive as very large domains must be used to form an accurate solution. Instead, we simulate the far-field condition using the Dirichlet-to-Neumann (DtN) method \cite{givoli}. This method is implemented by introducing a spherical artificial boundary, $R$, which is larger than the radius of the interface, i.e.\ $R > s(\theta, t)$.   Outside of this boundary
\begin{subequations}
	\begin{align}
	& \mbox{in}\quad r>R :  & \nabla^2u &= 0, \\
	& \mbox{on}\quad r=R: & u &= f(\theta),  \\
	& \mbox{as}\quad r \to \infty: & u &\sim 1,
	\end{align}
\end{subequations}
holds, where $f(\theta)$ is an unknown function. This problem can be solved exactly via separation of variables giving
\begin{equation} \label{eq:ExactBC}
u(r,\theta,t) = 1 + (c_0 - 1)\frac{R}{r} + \sum_{n=1}^{\infty} c_n \left(  \frac{R}{r} \right) ^{n+1} P_n(\cos\theta),
\end{equation}
where
\begin{equation} \label{eq:CoeffInt}
c_n = \frac{2n + 1}{2} \int_{0}^{\pi} f(\theta) P_n(\cos\theta) \sin\theta \textnormal{d} \theta
\end{equation}
and $P_n$ denotes the $n$th Legendre polynomial.  Matching this outer solution with the inner numerical solution on the artificial boundary $R$ provides the necessary Neumann boundary conditions for the numerical scheme.  By taking the derivative of \eqref{eq:ExactBC} with respect to $r$ at $r = R$ and evaluating \eqref{eq:CoeffInt} using the trapezoidal rule, the finite difference stencil for the radial derivatives is updated with
\begin{equation}
\begin{split}
\frac{u_{i-1,j} - 2u_{i,j} + u_{i+1,j}}{\Delta r^2} &+ \frac{2}{r_{i,j}} \frac{u_{i+1,j}-u_{i-1,j}}{2\Delta r} \to \\ &\frac{2(u_{i-1,j} - u_{i,j})}{\Delta r^2} + 2 \left( \frac{1}{\Delta r} + \frac{1}{R} \right)f'(\theta_j),
\end{split}
\end{equation}
where
\begin{equation}
f'(\theta_j) = \frac{1}{R} - \frac{(n+1)(\Delta \theta)}{R} \sum_{k=1}^{m-1} w_{j,k} u(R, \theta_k, t),
\end{equation}
and
\begin{equation}
w_{j,k} = \sum_{n=0}^{\infty} (n+1) P_n(\cos \theta_j)P_n(\cos \theta_k)\sin \theta_k. \label{eq:DtNSeries}
\end{equation}
From a practical perspective, we cannot, of course, evaluate the series in \eqref{eq:DtNSeries} using an infinite number of terms, but have found that using 10 terms gives sufficient accuracy. Furthermore, it is a straightforward exercise to use the DtN method for other types of far-field boundary conditions such as flux condition for fluid flow whereby $\partial u / \partial r \sim 1/r^2$ as $r \to \infty$.

\end{document}